\def\preprint{1}                
\def\comment#1{}
\if\preprint1
        \documentclass[useAMS,usenatbib]{mn2e}
      \usepackage{natbib}
        \usepackage{times}
        \usepackage{graphicx}
         \usepackage{url}
        \usepackage{calc}
        \usepackage{amssymb,amsmath}
        \usepackage{rotating}
         \usepackage{lscape}
         \bibpunct{(}{)}{;}{a}{}{,} 
         \usepackage{longtable}
\else
        \documentstyle[astrop-bib,referee,times]{mn2e}
        \newcommand{\includegraphics}[1]{}
        
\fi

\def\oversim#1#2{\lower0.5pt\vbox{\baselineskip0pt \lineskip-0.5pt
     \ialign{$\mathsurround0pt #1\hfil##\hfil$\crcr#2\crcr\sim\crcr}}}

\title[IPHAS Galactic supernova remnants]{New Galactic supernova remnants discovered with IPHAS}
\author[L. Sabin et al.]{L. Sabin$^{1,2}$\thanks{E-mail:lsabin@astro.iam.udg.mx (LS)}, Q.A. Parker$^{3,4,5}$, M.E. Contreras$^{1}$,  L. Olgu\'{i}n$^{6}$, D.J. Frew$^{3,4}$, M. Stupar$^{3,4,5}$, 
\newauthor R. V\'{a}zquez$^{1}$, N.J. Wright$^{7}$,  R.L.M Corradi$^{8,9}$, R.A.H Morris$^{10}$\\
$^{1}$Instituto de Astronom\'{i}a, Universidad Nacional Aut\'{o}noma de M\'{e}xico, Apdo. Postal 877, 22800 Ensenada, B. C, Mexico.\\
$^{2}${Instituto de Astonom{\'i}a y Meteorolog{\'i}a, Departamento de F{\'i}sica, CUCEI, Universidad de Guadalajara, Av. Vallarta 2602, C.P. 44130, Guadalajara, Jal., Mexico}\\
$^{3}$Macquarie University Research Centre in Astronomy, Astrophysics \& Astrophotonics, Sydney, NSW 2109, Australia\\
$^{4}$Department of Physics and Astronomy, Macquarie University, Sydney, NSW 2109, Australia\\
$^{5}$Australian Astronomical Observatory, PO Box 296, Epping, NSW 1710, Australia\\
$^{6}$Depto. de Investigaci\'{o}n en F\'{i}sica, Universidad de Sonora, Blvd. Rosales Esq. L. D. Colosio, Edif. 3H, 83190 Hermosillo, Sonora, Mexico\\
$^{7}$Harvard-Smithsonian Center for Astrophysics, 60 Garden Street, Cambridge, MA 02138, USA\\
$^{8}$Instituto de Astrof\'{i}sica de Canarias, E-38200 Tenerife, Spain \\ 
$^{9}$Departamento de Astrof\'{i}sica, Universidad de La Laguna, E-38200 Tenerife, Spain \\
$^{10}$Astrophysics Group, Department of Physics, Bristol University, Tyndall Avenue, Bristol, BS8 1TL, UK \\
}
\begin{document}

\date{Received 2012 June 11. Accepted 2013 January 25. 
}

\pagerange{\pageref{firstpage}--\pageref{lastpage}} \pubyear{2012}

\maketitle

\label{firstpage}

\begin{abstract}
 As part of a systematic search programme of a 10-degree wide strip of the Northern Galactic plane we present preliminary evidence for the discovery of four (and possibly five) new supernova remnants (SNRs). The pilot search area covered the 19--20~hour right ascension zone sampling from +20 to +55 degrees in declination using binned mosaic images from the INT Photometric H$\alpha$ Survey (IPHAS). The optical identification of the candidate SNRs  was based mainly on their filamentary and arc-like emission morphologies, their apparently coherent, even if fractured structure and clear disconnection from any diffuse neighbouring HII region type nebulosity. Follow-up optical spectroscopy was undertaken, sampling carefully across prominent features of these faint sources. The resulting spectra revealed typical emission line ratios for shock excited nebulae which are characteristic of SNRs, which, along with the latest  diagnostic diagrams, strongly support the likely SNR nature of these sources: G038.7-1.3 (IPHASX J190640.5+042819); G067.6+0.9 (IPHASX J195744.9+305306); G066.0-0.0 (IPHASX J195749.2+290259)  and G065.8-0.5 (IPHASX J195920.4+283740). A fifth possible younger, higher density nebula SNR candidate, G067.8+0.5 (IPHASX J200002.4+305035)  was discovered $\sim$5~arcmins to the west of IPHASX J195744.9+305306, and warrants further study. A multi-wavelength cross-check from available archived data in the regions of these candidates was also performed with a focus on possible radio counterparts. A close positional match between previously unrecognised radio structures at several frequencies and across various components of the 
H$\alpha$ optical image data was found for all SNR candidates. This lends further direct support for the SNR nature of  these objects.  Evolved SNRs may have very weak and/or highly fragmented radio emission which could explain why they had not been previously recognised but the association becomes clear in combination with the optical emission.   

\end{abstract}


\section{Introduction}
Supernova remnants (SNRs) are among the major contributors to the chemical regeneration of the Galaxy.  Indeed, following the explosion of their high mass progenitor stars in the case  of core-collapse supernovae ($\sim$80\% of all supernovae) or the Type-Ia binary route ($\sim$20-25\% of all supernovae, \citealt{Bazin2009}), they enrich the interstellar medium (ISM) in heavy elements (such as oxygen, iron and nickel) and other nucleosynthesis products. These ejecta play an important role in  the observed chemical composition of the local ISM and the stellar systems that may subsequently be formed from it. SNRs are also responsible for the release of large amounts of energy, generating shock waves that highlight or ``shape" the surrounding inhomogeneous ISM with the concurrent production of hot, dense and bright ionised zones. The detection and study of those rapidly evolving remnants (a few thousand years old) are crucial for the understanding not only of stellar evolution of  high mass stars in the case of core-collapse supernova but also, more globally, for the understanding of Galactic abundance gradients, kinematics and chemistry.

The most recent and complete compilation of SNRs in the Galaxy has been assembled by \citet{Green2009} which currently comprises 274 recognised and verified objects. However, according to \citet{Tammann1994} we should expect around 1000 SNRs to be visible at any one time in our Galaxy. This discrepancy could be accounted for by the lack (and difficulty) of detection at both ends of SNR evolution when either the SNR is too compact/distant to be seen or when it is on the verge of completely dissipating into the ISM.
The majority of the SNRs have been historically detected via radio observations and the subsequent determination of the non-thermal nature of the emission where the typically observed negative spectral index implies synchrotron emission. More recently optical detections of new Galactic SNRs have been possible (e.g. Stupar et al. 2008) even in the absence of previous radio evidence due to the advent of high resolution, high sensitivity H$\alpha$ surveys such as those described by Parker et al. (2005) and Drew et al. (2005).
 
\citet{Green2004} highlighted the most important radio surveys traditionally used for uncovering SNRs. These include  the Effelsberg 2.7~GHz survey \citep{Furst1990} with a resolution of $\simeq$4.3 arcmin, the Sydney University Molonglo Sky Survey (SUMSS) at 843~MHz  \citep{Bock1999}  with a resolution of 43$\times$43~arcseconds cosec $\delta$,  the Parkes-MIT-NRAO (PMN) radio continuum survey at 4.85~GHz \citep{Griffith1993} with a resolution of 
$\sim5$~arcmin for the Southern survey and $\sim$3~arcmin for the Northern counterpart undertaken at the same frequency on the Greenbank telescope (GB87; \citealt{Condon1994}) and finally the NRAO VLA Sky Survey (NVSS) at 1.4~GHz \citep{Condon1998} with a resolution of 45~arcseconds. These surveys are able to distinguish SNRs over a variety of angular scales depending on the respective beam sizes. For a review of SNRs in the PMN survey see \citet*{Stupar2007b}. 

Small, and therefore young (or distant), SNRs may not be optically recognised due to the presence of highly variable extinction at low Galactic latitudes where most SNRs are expected to reside. Other kinds of SNRs that are likely to be missed are those highly evolved SNRs with low surface brightness and large angular size which are usually also highly fragmented. Indeed, radio surveys are generally not very sensitive to extremely evolved SNRs (e.g. \citealt{Stupar2011a}, \citealt{Stupar2008}) where the detected radio fragments are often unrecognised. Furthermore, in the optical regime, the typically low emission levels from Galactic SNRs make them difficult to identify while confusion with the ubiquitous HII regions is also a problem. These selection effects have resulted in a serious bias in the global study of SNRs and their properties. 

As part of a systematic effort to help address this we are taking advantage of the recent, deep narrowband INT Photometric
H$\alpha$ Survey of the Northern Galactic plane (IPHAS; \citealt{Drew2005}, \citealt{Solares08}) to look for ``previously hidden" but optically detectable SNRs in a similar way to those uncovered by \citet{Stupar2008} in the South.
IPHAS, which started in 2003, is now essentially complete. It used a wide-field CCD camera with $r$, $i$ and H$\alpha$ filters to provide an accurate photometric survey.  IPHAS covers 1800 deg$^2$  of the northern plane across a $\pm$5~degree latitude strip and was initiated  as the direct counterpart to the AAO/UKST H$\alpha$ survey of the Southern Galactic plane \citep{Parker2005}. The survey has a nominal resolution of 0.33 arcsec.pix$^{-1}$ and the narrow-band component allows the detection of extended optical emission with an H$\alpha$ surface brightness down to $\simeq$ 2$\times$10$^{-17}$ erg cm$^{-2}$ s$^{-1}$ arcsec$^{-2}$ ($\simeq$3 Rayleighs). This makes the survey sensitive to evolved, low surface brightness nebulae such as planetary nebulae (PNe), including those interacting with the ISM  \citep{Sabin2010,Wareing2006}, symbiotic stars \citep{Corradi2010}, Wolf-Rayet (WR) shells \citep{Stock2010}, Herbig-Haro objects, proplyd-like objects \citep{Wright2012}, and other kinds of resolved nebulae as well as SNRs.

This is not the first time a deep, narrow-band optical survey has been used to search for new SNRs as previously mentioned. 
\citet{Stupar2007a,Stupar2008,Stupar2011b} presented the discovery of a significant number of new optically identified and confirmed SNRs based on data from Southern Galactic plane H$\alpha$ Survey \citep{Parker2005}. The authors also published a catalogue of newly uncovered optical counterparts for 24 known Galactic SNRs previously identified only from radio or X-ray observations \citep{Stupar2011a}. \citet{Boumis2005,Boumis2008,Boumis2009} also presented the first optical detections and spectroscopic confirmation for a number of SNRs based on CCD data. While also in the northern sky \citet{Fesen2010} discovered optical filaments for SNR G159.6+7.3 which had no previous radio detection. 

In this paper we report the preliminary results of a systematic search for new SNRs from the IPHAS survey. Radio observations (across a large frequency range) are widely used to identify and image SNRs so in addition to the character of the ionised optical emission presented, the detection of radio emission from the expanding SNR shell is another important tracer, especially if it is possible to measure a negative spectral index to indicate synchrotron radiation.  We examined archival data from the Westerbrock Northern Sky Survey (WENSS) at 325~MHz \citep{Rengelink1997}, the NRAO VLA Sky Survey (NVSS) at 1.4~GHz \citep{Condon1998} and the Green Bank (87GB) 4.85~GHz survey \citep{Gregory1996}  to look for radio counterparts to our IPHAS targets. Furthermore, the high energy released by the initial supernova explosion (typically about 10$^{51}$ ergs following \citealt{Burke2002}) results in the generation of X-ray emission which can be detected by X-ray space telescopes such as the previous ROSAT and current Chandra and XMM Newton  space observatories. Archival data from these surveys have also been interrogated to look for possible X-ray counterparts, both diffuse and compact. A search for infrared counterparts was also undertaken using most of the available IR surveys (e.g. WISE, GLIMPSE) but this did not return any convincing correspondence.

 The paper is organised as follows. In \S~2 we present the new IPHAS optical detections of the SNR candidates. In \S~3 we present the spectroscopic follow-up of the candidates. In \S~4 some multi-wavelength, principally radio counterparts to the optical images are discussed, while in \S~5 we present a discussion and our conclusions.

\section{IPHAS detections of new SNR candidates}
This initial search for new, optically detectable SNRs concentrated on the right ascension range from 19h--20h. This is the first major IPHAS zone which has been systematically and carefully scanned for new emission nebulae of all kinds, including PNe and HII regions. This search formed a major component of the thesis of  \citealt{sabinthesis}. A separate paper dedicated to the first 100 confirmed PNe resulting from this search is also in train (Sabin et al. 2013, in prep.). Note that the full IPHAS sky area (1800 deg$^{2}$) has been divided into 2 deg$^2$ H$\alpha$-r (continuum removed) image mosaics that facilitate detection and recognition of extended emission nebulae. We used mosaic scale factors of $15 \times 15$ pixels (5~arcsec) and $5 \times 5$ pixels (1.7~arcsec). The first binning level was chosen to enable the detection of low surface brightness objects (down to the IPHAS limit) \citep{Sabin2010} over large angular scales while the second was used to detect intermediate size nebulae, i.e smaller than $\simeq$15--20 arcsec in diameter.

 The SNR candidates were selected based on their optical morphological characteristics. They were considered as plausible candidates if they could be identified as  detached but coherent filamentary structures or as discrete round shells or arcs in the IPHAS H$\alpha$-r mosaic images. The H$\alpha$ Balmer emission line is an excellent tracer of ionised gas while the continuum subtracted images enhance the contrast even further. Already catalogued sources, including known SNRs are also scrutinised as additional, previously unrecognised emission components to these sources may become evident. For example most known SNRs were uncovered only from radio data but we can now also reveal optical counterparts to many of these remnants for the first time \citep{Stupar2011a}.

Spectroscopic validation of newly identified optical components of perhaps widely dispersed SNR candidates is important for their proper identification but this can be difficult to realize in crowded zones such as the Galactic centre where several SNRs (and HII regions)  can overlap \citep[their figure 3]{Green2009}.  Our careful search over the preliminary RA zone resulted in a set of four or possibly even five candidate SNRs being uncovered. These are described individually below with summary details presented in Table\ref{listSNR} and named according to the nomenclature adopted for all IPHAS extended objects (IPHASX JHHMMSS.s$\pm$DDMMSS).

 Our first candidate SNR IPHASX J190640.5+042819 was selected due to its well defined semi-circular emission of radius $\sim$9.3\arcmin $\times$15.9\arcmin\ together with some elongated tails of filamentary features particularly to the South. These morphological structures classified it as a good SNR candidate (Fig.\ref{SNR1906}--Top). 

The second candidate is IPHASX J195744.9+305306  (Fig.\ref{SNR195744.9}--Top) which has the appearance of a fractured oval ring with some internal emission spurs emanating from the Eastern and Northern edges. It has a projected angular radii of $\sim$ 22.6\arcmin$\times$24.6\arcmin. There are several  HII regions in the general vicinity of this candidate SNR, including Sh~2-98 and Sh~2-97 which are found within $\sim$0.5 deg but these are well separated from the SNR candidate. However, the most northerly faint rim of the SNR appears somewhat detached from the main shell and comes within 4 arcminutes of the outer rim of the circular HII region directly to the North so some overlapping of  emission here cannot be completely ruled out.  There is also a known SNRs  G067.7+01.8, located $\sim$1~degree west. Clearly, this general area  hosts some large nebulae.  Nevertheless, the main structure remains clearly detached from any of the surrounding HII regions and we can consider IPHASX J195744.9+305306 as a single object.  

A coherent, completely separate, elliptical emission structure was also found $\sim$5\arcmin\ to the east of IPHASX J195744.9+305306 (Fig.\ref{PointingD}--Top) which we label IPHASX J200002.4+305035 and was included in our investigation as it could be a younger, more compact SNR or perhaps even an evolved PN. 

We also identified IPHASX J195749.2+290259 (Fig.~\ref{195749}) which has a filamentary partial shell morphology with a bright emission arc to the North. A fainter apparent morphological duplication of this structure $\sim$5\arcmin\ to the North-West is also evident which must be related given their strong similarity in appearance.  Such similar morphological repetition is also seen in parts of the well known Vela SNR. We cannot probe deeper into  a detailed explanation at this stage as we lack the data  to fully establish the nature of the''duplicated" structure which may represent some evidence of symmetrical mass ejection. Further study is needed but this is beyond the scope of this paper. The structure of IPHASX J195749.2+290259 is again typical of optical (and radio) SNRs and we note the  similarity between this object and the well known SNR IC~443 which also exhibits faint filament (G189.1+3.0; see \citealt{Fesen1984}). 

The last selected SNR candidate is IPHASX J195920.4+283740 which reveals an elongated and closed shell structure at faint isophotes over a region of radius 186$\times$312~arcseconds (Fig.\ref{SNR1959}--Top). The nebula is globally very faint except for the enhanced edge seen in the Eastern side.

\begin{table}
\begin{center}
\begin{small}
\caption{\label{listSNR} {Estimated central positions and angular semi minor and major axes for the new SNR candidates found via visual inspection of the IPHAS mosaics in the 19--20~hour RA zone. The RA/DEC positions are J2000.}}
\begin{tabular}{|c|l|c|clc} 
\hline
 IPHASX ID  &   Galactic ID  &RA (J2000) &  DEC (J2000)          &  Angular size \\
                & & h~m~s & $^{0}$~'~'' & (arcmin) \\
\hline
 J190640.5+042819 & G038.7-1.3 & 19:06:40.5& +04:28:18.8 & 18.6$\times$31.8 \\
 J195744.9+305306 & G067.6+0.9 & 19:57:44.9 & +30:53:05.6 & 45.2$\times$49.2\\
 J195749.2+290259 & G066.0-0.0 & 19:57:49.2 & +29:02:58.8 & 24.6$\times$31.0  \\
 J195920.4+283740 & G065.8-0.5 & 19:59:20.4 & +28:37:40.0 & 6.2$\times$10.4\\
 J200002.4+305035 &  G067.8+0.5 & 20:00:02.4 & +30:50:35.0 & 7.2$\times$5.4\\
\hline
\end{tabular}
\end{small}
\end{center}
\end{table}
\begin{table}

\begin{small}
\caption{\label{specSNR} {Summary details of the spectroscopic observations of the SNR candidates. The formal wavelength ranges of the spectra are included}}
\begin{tabular}{lllcl}
\hline
 IPHASX ID   & Telescope  &  Date   & $\lambda$ coverage & Exposure\\
                       &                &            & {\AA}                      & (seconds)\\
 \hline
 J190640.5+042819 & SPM 2.1m &  07/06/2010 & $\simeq$4330--7530 & 1800 \\
 J195744.9+305306 & SPM 2.1m &  21/08/2011 &  $\simeq$4330--7530& 1800\\
 J195749.2+290259 & SPM 2.1m &  22/08/2011 &  $\simeq$4330--7530& 1800 \\
 J195920.4+283740 & INT 2.5m   & 18/04/2008 &  $\simeq$3500--8960&  $3\times$1200 \\
 J200002.4+305035 & SPM 2.1m & 22/08/2011 & $\simeq$4330--7530  & 1800  \\
\hline
\end{tabular}
\end{small}
\end{table}
\section{Spectroscopic observations and confirmatory results}
As part of our general programme of IPHAS follow-up spectroscopy, we obtained low resolution confirmatory spectra of these candidate SNRs.
 Our first SNR spectrum was of IPHASX J195920.4+283740 taken on the 2.5m Isaac Newton Telescope on La Palma (Canary Islands) in April 2008  while the remaining candidates were observed using the 2.1-m telescope at the San Pedro Martir Observatory (OAN-SPM\footnote{The Observatorio Astron\'omico Nacional at the Sierra de San Pedro M\'artir (OAN-SPM) is a national facility operated by the Instituto de Astronom\'{\i}a of the Universidad Nacional Aut\'onoma de M\'exico.}) between June 2010 and August 2011. A summary of these spectroscopic observations are given in Table \ref{specSNR} while the spectra are presented in Fig.\ref{Spectra} and the resulting emission line measurements are shown in Table \ref{Em}.

On the 2.5m INT we used the low resolution spectrograph IDS with the EEV10 detector and R300V grating. This covers the range $\simeq$ 3500 {\AA}-8960 {\AA} with a resolution of $\simeq$ 1.8 {\AA} though the data below 3600 {\AA} is of no value due to the atmospheric cut-off.  The slit length was 3.3 arcminutes and the seeing during the observations varied between $\sim$1.8 and $\sim$2.2 arcseconds. The exposure time for the observation of  IPHASX J195920.4+283740, which is a relatively faint nebula was set to 3$\times$1200s on a portion of its brighter Eastern edge. We note the data blueward of H$\beta$ in the spectrum did not show any emission lines due to the low S/N.\\
For the remaining 2.1-m San Pedro Martir telescope observations we used a Boller $\&$ Chivens spectrograph and  Thomson $\&$ Marconi  CCDs (2048$\times$2048 pixels of size $14\micron$). An East-West slit orientation was adopted with the 5 arcminute length slit carefully positioned on the most prominent nebular arcs or filaments to give better S/N but also chosen to allow sufficient sky for background subtraction. In cases where, despite our best efforts, it was not possible to extract sufficient decent sky we also took background spectra in relatively''blank" areas located in the immediate surroundings of the nebula. This was particularly the case for IPHASX J195744.9+305306 (B \& C positions). 
The observations were obtained under varying seeing conditions typically around 2~arcseconds and were made with the 400 line/mm grating. This gave a spectral coverage of $\simeq$ 4330 {\AA}--7530 {\AA} with a resolution of $\simeq$ 5 \AA\, sufficient to encompass most optical emission lines of interest.  The data blueward of H$\beta$ in these spectra did not exhibit any emission lines  due to the low S/N and extinction. The data reduction of all spectra was done with the usual IRAF routines. The extinction correction applied to our spectra was performed using the Fitzpatrick and Massa extinction curve \citep{Fitzpatrick2007} for $R_{V}$ = 3.1. 

In order to derive the errors on the measurements we consider mainly two sources of error. The first is related to the CCD readout noise and photon noise relative to the sky and objects. This error is derived via the estimation of the variance ($\sigma^{2}$) at each point of the spectrum (i.e. at each wavelength). The "variance spectrum" obtained is processed the same way as for regular data spectra in IRAF assuming a non-correlated error approximation \citep{Taylor1997}. Therefore the statistical errors corresponding to each spectral line are extracted and registered. The second source of error we considered here is linked to the flux calibration and the data were processed following \citet{Viironen2011}. 

The usual H$\alpha$,  [N\,{\sc ii}]$\lambda${$\lambda$}6548,6583 and  [S\,{\sc ii}]$\lambda${$\lambda$}6716,6731 emission lines were used to identify the nature of the nebulae based on the criteria defined by \citet{Fesen1985}[hereafter F85] and refined most recently by \citet{Frew10}. The most conspicuous optical features for the discrimination of SNRs are the relative strength of  [S\,{\sc ii}] lines compared to H$\alpha$. Indeed a large ( [S\,{\sc ii}]6716+ [S\,{\sc ii}]6731)/H$\alpha$ ratio $>\,$0.4--0.5 (see F85) is a key probe for the occurrence of the shock characteristics of the turbulent and energetic environment of SNRs which is not generally found in HII regions or PNe.

The  [S\,{\sc ii}] ratio is used as a diagnostic to determine electron densities in ionized plasmas and for most SNRs this ratio is likely to be in the low density limit with a ratio of $\sim$1.4. In addition, the presence of relatively strong  [N\,{\sc ii}] lines (with 0.5\,$<$( [N\,{\sc ii}]6548+ [N\,{\sc ii}]6583)/H$\alpha$\,{$<$}\,1.5 following F85) as well as the presence of moderate [OI] emission ($\lambda${$\lambda$}6300,6363) are also good indicators of a likely SNR nature. However, the latter can cause problems in low-resolution spectra as these lines are also quite strong in the night sky making their deconvolution from their bright sky line equivalents very difficult unless there is a strong velocity component in the SNR [OI] lines. This is especially germane if the extended nebulosity extends along the entire length of the slit. Our main limitation though is the inherent faintness of the targets even in H$\alpha$, which also explains their non-detection in previous optical surveys. An immediate consequence of such faint nebulosity is the inability to easily detect other common diagnostic emission species such as the Fe lines \citep{Fesen1996}. 

Another issue related to the faintness of our targets is the discrimination between true SNRs and other morphologically similar nebulae such as Wolf-Rayet (WR) shells, symmetric HII regions or even low excitation zones in PNe, as these nebulae can exhibit similar emission-line ratios in some cases. An overall body of evidence is used to aid in the identification of likely SNRs and to remove contaminants. This comprises the optical morphology, spectroscopic signature, local environment, presence of X-ray sources internal to the nebulae and any corroborating radio emission.   While other shock-excited nebulae such as supershells \citep{Hunter1994,Skelton1999} and giant outflows associated with Herbig-Haro objects \citep{Reipurth1997,Mader1999} can have similar emission line-ratios to SNRs, the small (less than the degree) angular sizes of  our candidates, their filamentary morphologies and the lack of star-forming activity in their vicinities strongly militates against these alternative interpretations.

\subsection{New emission line diagnostic diagrams}

 The ability to separate out different classes of astrophysical object according to their observed emission-lines ratios is well established, e.g. the well-known \citet*{Sabbadin77} `SMB' H$\alpha$/{[N \textsc{ii}]} versus H$\alpha$/{[S \textsc{ii}]} diagnostic diagram or the \citet*{Baldwin1981} `BPT' {[O\,\textsc{iii}]}$\lambda$5007/H$\beta$ versus {[S \textsc{ii}]}/H$\alpha$ diagnostic diagram. We have been refining and updating these diagnostic diagrams using carefully vetted spectroscopic data from the literature, supplemented with our own flux-calibrated spectroscopy for a significant sample of objects of various kinds.  The new figures presented here for the first time are better populated versions of the diagrams presented by \citet{Frew10}.  These new plots include many more Galactic SNRs and other types of nebulae than previous efforts.  By only utilising the most reliable data and identifications, a clearer picture of the loci of different object types is established compared to what has been previously available.

In this paper we therefore base our SNR spectroscopic confirmations on their observed strong emission line-ratios used in conjunction with these powerful, updated diagnostic diagrams created by one of us (DJF), based on an extensive database of spectroscopic measurements.  The line-ratio data for the Galactic SNRs was taken from our own unpublished spectroscopy as well as numerous literature sources, with notable contributions from \citet{Fes1980}, \citet{Fes1982b,Fesen1985}, \citet{Blair1991}, and \citet[and references therein]{Boumis2005}.  The LMC SNR flux data is mostly from \citet[and references therein]{Payne2008} and the line fluxes for Galactic PNe are taken from the sources summarised in \citet{Frew2012}.  The HII region data are mostly taken from the references given in \citet{Frew10}, supplemented with unpublished data utilized by \citet{Boissay2012}.  The labelled domains include approximately 95\% of the objects of that class listed in our database.  A fuller description of the construction and content of these new diagnostic diagrams will be presented in their entirety in an upcoming paper (Frew et al., 2013, in preparation, F13 hereafter).  Our new sources are over-plotted on these figures which demonstrates clearly how they lie in the zones populated by confirmed SNRs (see Figure~\ref{DJF} and Figure~\ref{DJF2}). 

We emphasise that evolved Galactic SNRs have compositions dominated by swept-up ISM, and show a clear abundance gradient in the Galaxy (\citealt{Binette1982}; F85); the most metal-rich examples in the inner Galaxy tend to plot at the lower-left of the SNR domain in the H$\alpha$/{[N \textsc{ii}]} versus H$\alpha$/{[S \textsc{ii}]} diagram (Figure~\ref{DJF}). Since the IPHAS candidates also plot here, this suggests that they are metal rich, which is consistent with the Galactic sight-lines for these objects. At any plausible distance, they are located interior to the Solar Circle.  This diagnostic plot also illustrates how the Galactic objects tend to plot separately from the Magellanic Cloud remnants, a consequence of the lower nitrogen abundance in the latter  \citep{JMeaburn2010,Leonidaki2012}.  We also note that many of the youngest core-collapse Galactic SNRs (e.g. \citealt{Fes1982}; \citealt{Leibowitz1983}; \citealt{Winkler1985}) are chemically stratified, and show clear nitrogen overabundances in many of their filaments (\citet{Chevalier1978,Chevalier1979}; \citet{MacAlpine1994,MacAlpine1996}; \citealt{Isensee2012}).

The individual spectroscopic details of each IPHAS candidate are described in the following subsections.

\subsection{IPHASX J190640.5+042819} 
For this candidate (refer to the images in Figure~\ref{SNR1906}) our new optical spectrum gives ratios of  [N\,{\sc ii}]/H$\alpha$=1.48 (eliminating any possible confusion with a HII region) and  [S\,{\sc ii}]/H$\alpha$=1.15, clearly placing the ionised gas in the shock regime. Placing these emission line ratios in the F13 and F85 diagnostic diagrams, indicates the source lies squarely in the region occupied by SNRs. The forbidden oxygen emission lines [OI] $\lambda${$\lambda$}6300,6364 which are relatively strong in SNR  (contrary to HII regions and PNe) are also present and were measured with extinction-corrected fluxes of 34.1 and 34.5 respectively (for H$\beta$=100 and H$\alpha$=286). The [OI] $\lambda$6300 line flux should be $3\times$ the [OI] $\lambda$6364 flux according to the theoretical ratio of Einstein's A coefficients. The observed values betray sky subtraction problems due to the presence of the same oxygen lines in the sky-background that makes them difficult to extract properly as the nebular emission filled most of the slit. Although we proceeded with great care when removing the sky, there is clearly significant uncertainty in the flux measurement of these lines. IPHASX J190640.5+042819 is our first confirmed SNR which can then be named according to the standard Galactic SNR nomenclature (GD.d{$\pm$}D.d) as SNR G038.7-1.3. 

\subsection{IPHASX J195744.9+305306}
The large projected optical angular extent of IPHASX J195744.9+305306 ($\sim$24.6\arcmin$\times$22.6\arcmin -  see Figure~\ref{SNR195744.9}) led us to obtain several different pointings during the spectroscopic observations so as to obtain a more representative sample of spectra across different possible components of this large source. Those are noted as A, B and C in the top panel of Figure~\ref{SNR195744.9}. There is an additional pointing D which was made on the faint, oval shaped, apparently detached nebula $\sim$5\arcmin\ to the east of the main structure and approximately 8\arcmin$\times$6\arcmin\ in diameter (Figure~\ref{PointingD}). This turns out to be an interesting source in its own right and another possible SNR candidate (see below). 

Figure~\ref{DJF} indicates that the spectra from pointings B and particularly C provide line ratios compatible with an SNR origin with a clear signature of shock-heated emission with  [S\,{\sc ii}]/H$\alpha$ ratios of 0.57 and 0.99 respectively (Figure~\ref{DJF2}, Table \ref{Em})- see also the diagram by \citet{Phillips1999}. Pointing B also shows the [OI]$\lambda${$\lambda$}6300,6364 emission lines with respective dereddened values of 41 and 11 (scaled to H$\beta$=100), supporting the SNR classification (see Figure~\ref{Spectra} and Table \ref{Em}). However, the observed line ratios from the spectra for pointing A are closer to those expected for a HII region (Table \ref{Em}). Note that pointing `A' falls on a nebular condensation internal to the proposed main SNR shell and could simply be an unrelated compact HII region, as, unlike the two internal emission spurs to the North, this blob does not appear connected to the main arcuate shell. Furthermore, the spectral characteristics of this region (slit position A) are different to those obtained from slit positions B and C with instead a rather low  [S\,{\sc ii}]/H$\alpha$ ratio (0.24) which does not advocate for a shock regime and very different [NII]/H$\alpha$ ratios, adding further weight to its identification as a discrete HII region.  Nevertheless, given we have no independent estimate of the distances of these components we cannot categorically discard a link between this putative compact HII region and the surrounding main SNR shell. Additional information such as distance estimations and the kinematics of both structures are required to fully establish any possible connection.

The information brought by the extinction data would position the filaments/structures related to pointing A and B at relatively close distance from each other with respective $c$(H$\beta$) values of 0.82 and 0.86. Pointing C, with $c$(H$\beta$)=1.26, would then appear as a independent element but we cannot discard a link to IPHASX J195744.9+305306 since pointings `B' and `C' share the same SNR spectroscopic classification. 


 In summary SNR candidate IPHASX J195744.9+305306  falls in a  general area of HII regions including a probably compact HII region (pointing A) superposed on the main SNR shell. HII regions projected on SNRs are not unexpected and our combined imaging and spectroscopic data still provide strong evidence that  IPHASX J195744.9+305306 (G067.6+0.9) is a true SNR. 

\subsection{IPHASX J200002.4+305035}
The spectral characteristics exhibited from pointing D (object shown in Figure~\ref{PointingD}) including from its location in the F13 diagram, and a quite high but not conclusive  [S\,{\sc ii}]/H$\alpha$ ratio,  that this apparently coherent, oval nebulae may be an additional compact SNR which we tentatively denote as IPHASX J200002.4+305035 or G067.8+0.5.  It is located ~5\arcmin\ to the East of the more extensive nebulosity linked to IPHASX J195744.9+305306 and is clearly detached from it. The spectrograph slit was placed E-W across the middle of the Western enhanced edge of the optical nebula. The absence of the H$\beta$ line prevents us from deriving the extinction. Note that it is unrelated to the nearby compact (FWHM $\sim$1.62~arcseconds) PN candidate IPHAS~J195956.42+304823.8 (2MASS J19595642+3048238) reported by \citet{Viironen2009}. IPHASX J200002.4+305035 also shows the highest electron density of all the targets with n$_{e^{-}}$= 620$\pm$74~cm$^{-3}$ which may be an indication of youth. Further study of this emission nebula is required to clarify its nature but an SNR identification is not unwarranted at this stage from the optical imagery and spectroscopy alone.

\subsection{IPHASX J195749.2+290259}
The spectroscopic analysis of this candidate  shown in Figure~\ref{195749}, indicates a likely SNR identification (in this case the nebula is named G066.0-0.0) either using the diagnostic diagrams from F13   (Figure~\ref{DJF}) or the conditions fixed by F85. The observed diagnostic emission line ratios of  [S\,{\sc ii}]/H$\alpha$=0.83 (clearly in the shock regime following F85),  [N\,{\sc ii}]/H$\alpha$=1.9 (clearly eliminating a HII region identification) and  [S\,{\sc ii}]6717/6731=1.4 indicating an electron density close to the low density limit,  all fit well within the definition of an SNR following F85. We notice that, similarly to G067.6+0.9 (IPHASX J195744.9+305306: pointings A and C), we did not detect any clear [OI] emission (see Figure~\ref{Spectra}).

\subsection{IPHASX J195920.4+283740}
The nebula  presented in Figure~\ref{SNR1959}, is well constrained spectroscopically to fall within the SNR regime using the criteria of F85 ( [S\,{\sc ii}]/H$\alpha$=0.90 and  [N\,{\sc ii}]/H$\alpha$=1.25) and F13 (see Figure~\ref{DJF}).  However, a new helium rich emission point source possibly related to a WR star has been identified close to the enhanced bow-shaped eastern emission component by \citet{Corradi2010}, and denoted as IPHAS J195935.55+283830.3. Possible association of the nebulae with this unusual star would throw doubt on  its SNR nature  and might imply that the shell results, at least partly, to a past ejection from a massive star.  We note that the elongated and asymmetric filamentary morphology of IPHASX J195920.4+283740 does not show the ring-like structure(s) generally displayed by WR nebulae (\citealt{Stock2010}, \citealt{Marston1995}), which lends some support to an SNR classification.  However, the IPHAS nebula does not display the typical SNR filamentary morphology either.  The SMB log\,(H$\alpha$/ [N\,{\sc ii}]) versus log\,(H$\alpha$/ [S\,{\sc ii}]) diagnostic diagram after \citet{Frew10} is useful in discriminating SNR and WR shell emission-line signatures as they tend to occupy different loci in the diagram. In the case of IPHASX J195920.4+283740, the object plots well inside the SNR area and away from the WR shell / HII region domain. Nevertheless, the presence of a helium rich star so close to the enhanced eastern edge of the nebula is suspicious such that a non-SNR origin cannot be completely ruled out, despite the indicative emission-line ratios and corroborating radio signatures (see later).

\begin{figure}
\begin{center}
\hspace{-1.2cm}
{\includegraphics[height=7.3cm]{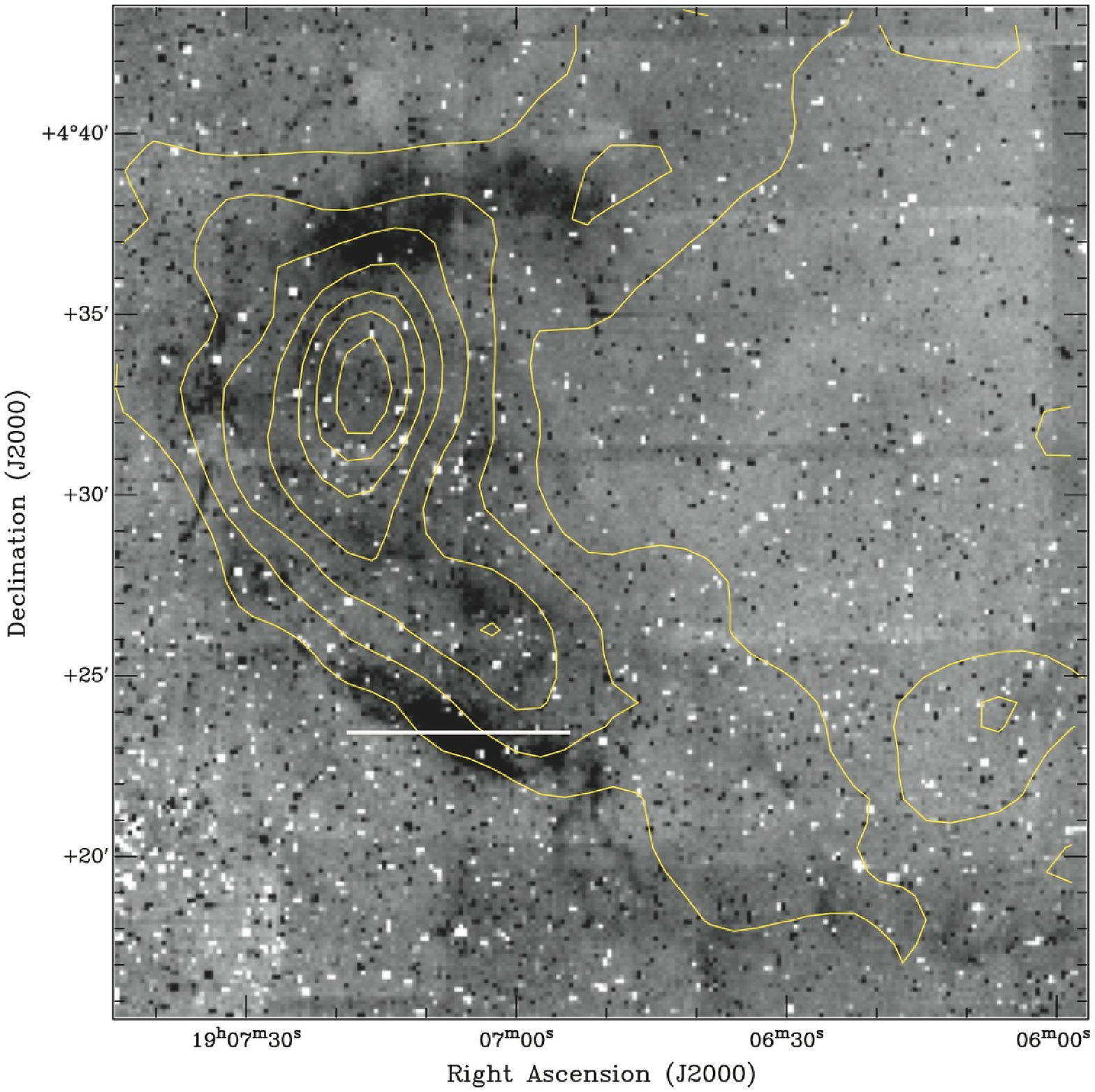}}
{\includegraphics[height=6.4cm]{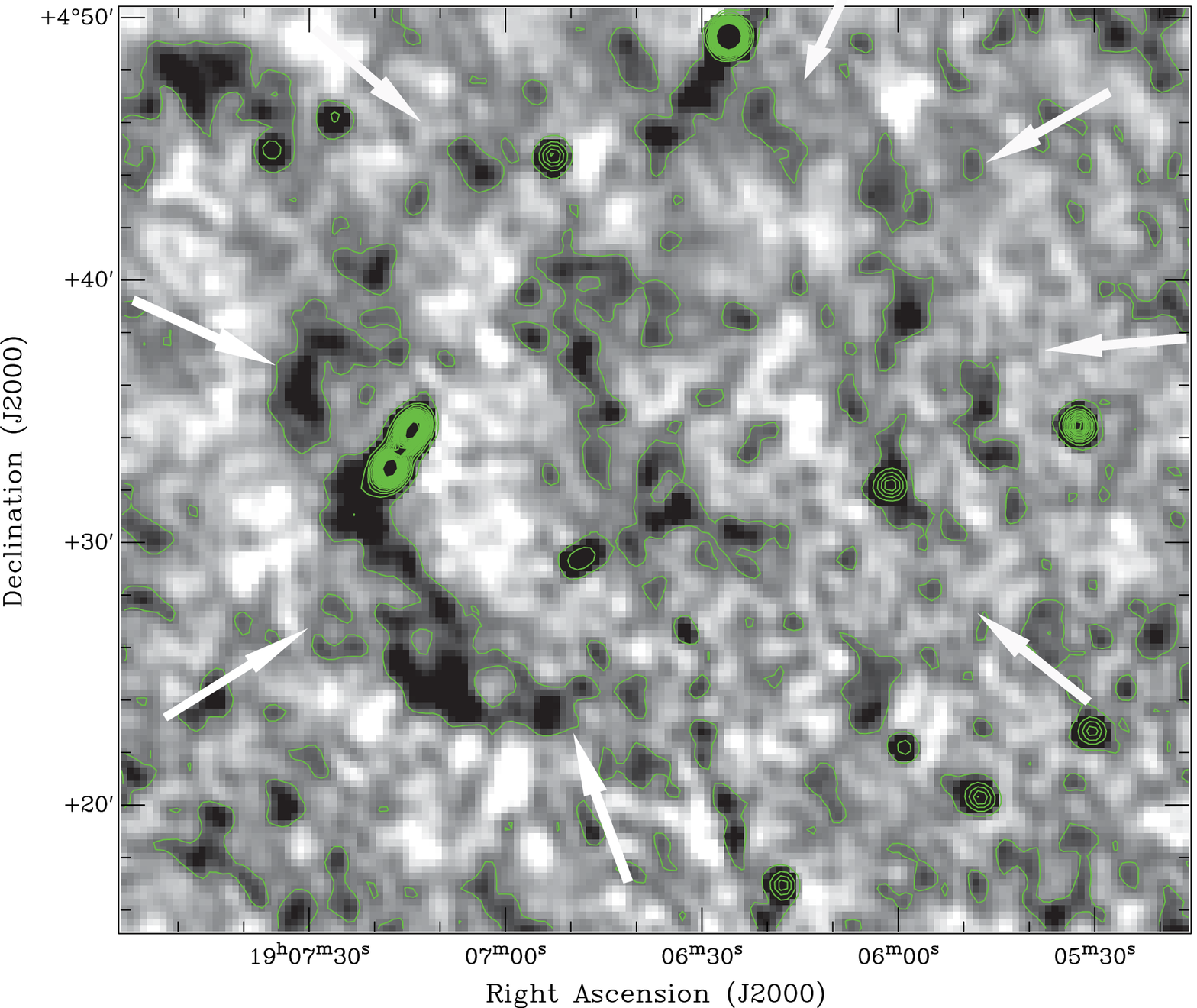}}
\caption{\label{SNR1906}  The top panel presents a binned (5~arcsecond/pixels) H$\alpha$ image of IPHASX J190640.5+042819 from the IPHAS mosaics overlaid with 6~cm contours from the 87 GB survey. Contours go from 0.01 to 0.14 Jy/beam in intervals of 0.02~Jy/beam. There is an excellent match between the  H$\alpha$  and the radio data. The peak flux has 90~mJy. The bottom image presents the NVSS 1.4 GHz (21~cm) radio image. The positional match with H$\alpha$ and the 6~cm emission is obvious on the eastern side. Due to the improved resolution of $\sim$45~arcseconds, the whole SNR shell (delimited by the white arrows) can be noticed which enables an updated SNR ID as G038.7-1.3 based on the geometric centre of the shell. The NVSS radio contours go from 0.001 to 0.05 Jy/beam in intervals of 0.005~Jy/beam. The position of the spectrograph slit is indicated.}
\end{center}
\end{figure}

\begin{figure}
\begin{center}
{\includegraphics[height=7.5cm]{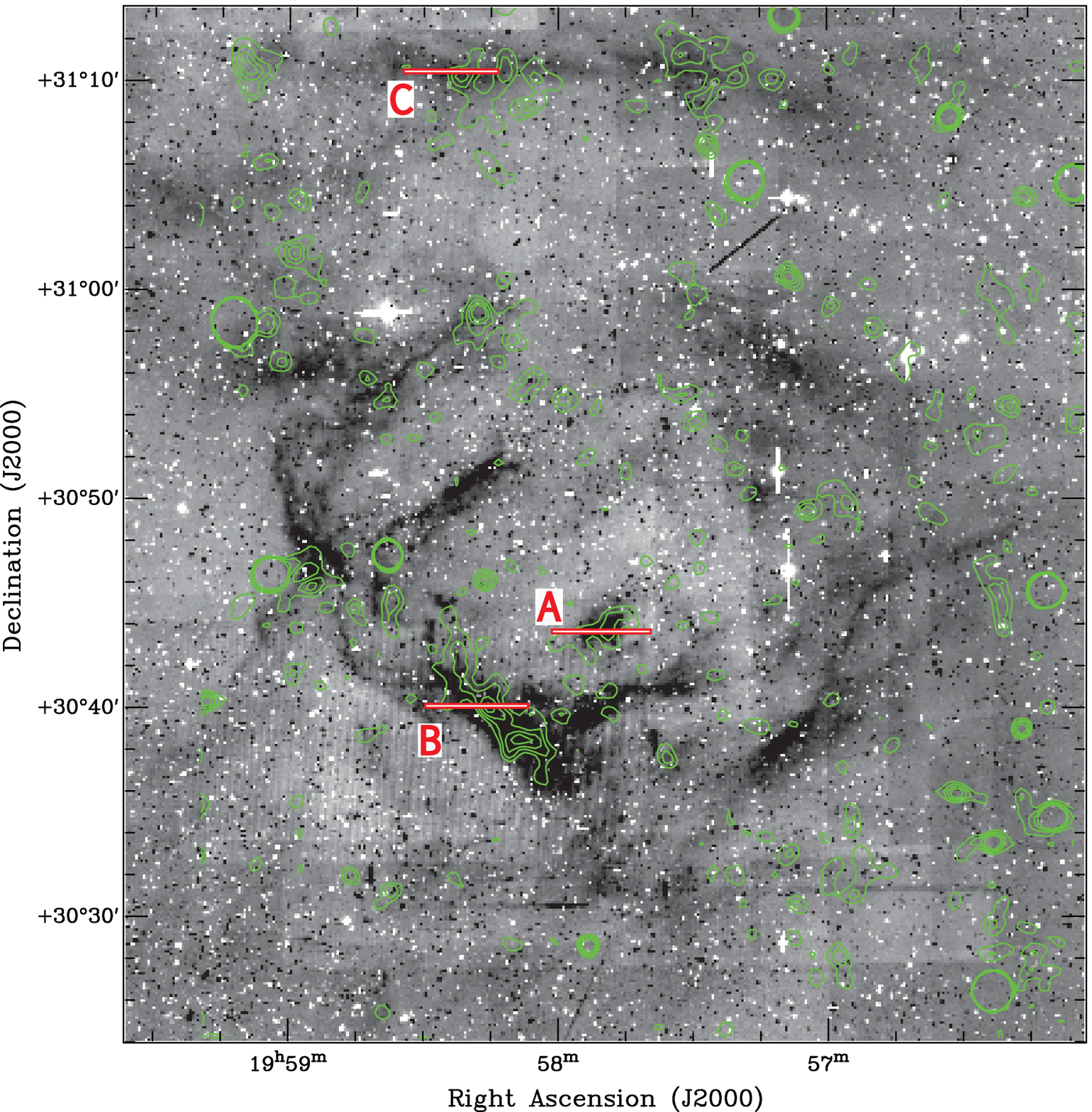}} 
{\includegraphics[height=7.5cm]{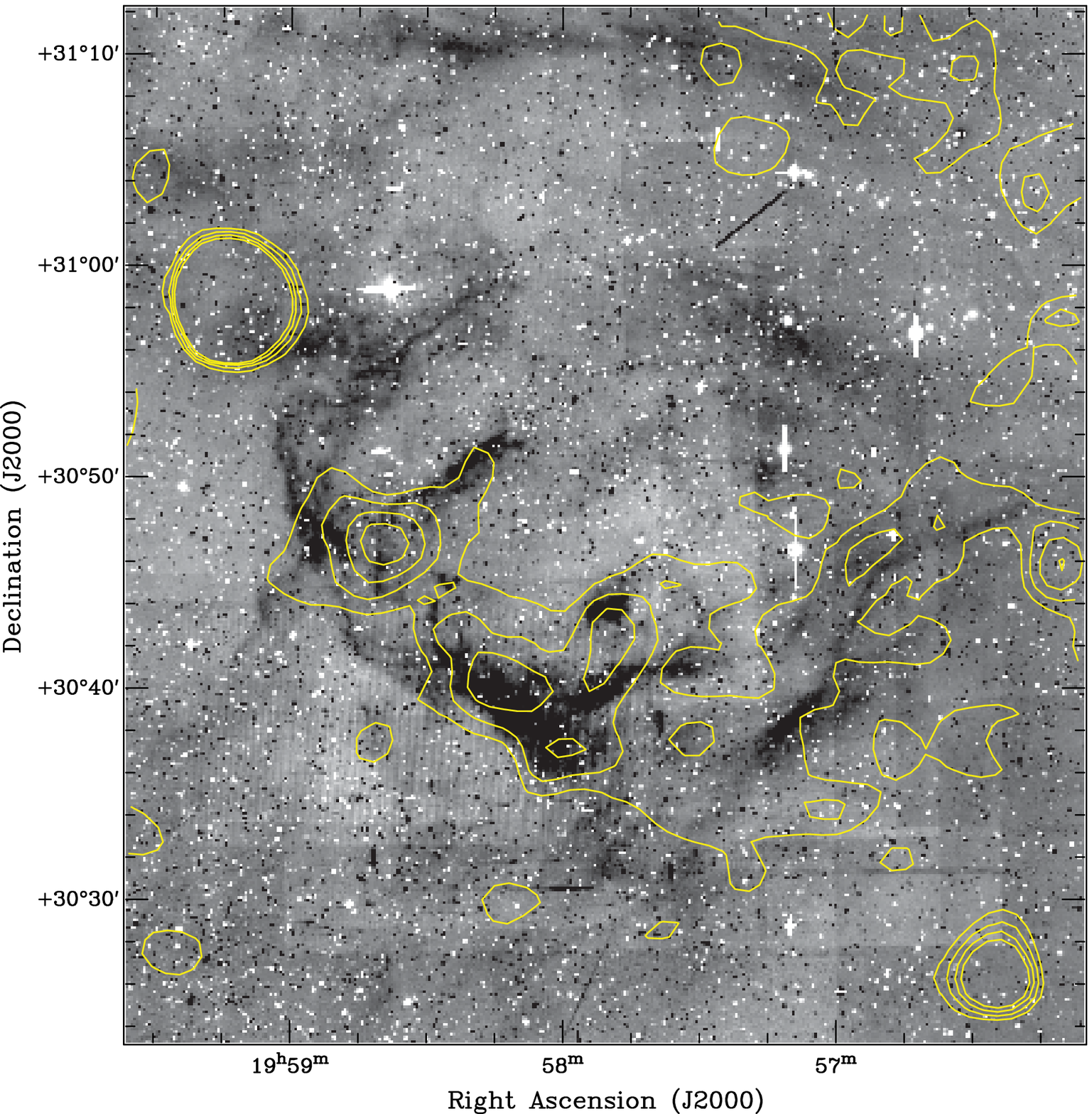}} 
\caption{\label{SNR195744.9} Top panel: The H$\alpha$-r image mosaic of IPHASX J195744.9+305306 with 15$\times$pixel binning overlaid with NVSS radio contours in green ranging from 0.0008 to 0.0027 Jy/beam. The red E-W bars indicate the location of the slits for the different spectral pointings A, B and C with lengths of 5\arcmin.  The nebula is in a generally complex environment which makes it harder to define the real limits of the new SNR in H$\alpha$ at the Northern fainter rim (sampled by slit C). The diagonal dark streak is an image artefact (possibly an asteroid trail). Bottom panel: As above but now with the lower resolution 6~cm 87 GB  radio survey data overlaid in yellow contours which range from 0.004 to 0.02 Jy/beam. As in the case of the 21~cm NVSS data the 6~cm  87 GB radio emission follows the edges of the SNR very closely and especially along the South East part of the nebula where the optical emission is strongest. The optical-radio overlap is not as complete as in the case of the SNR IPHASX J190640.5+042819. }
\end{center}
\end{figure}

 \begin{figure}
{\includegraphics[height=6cm]{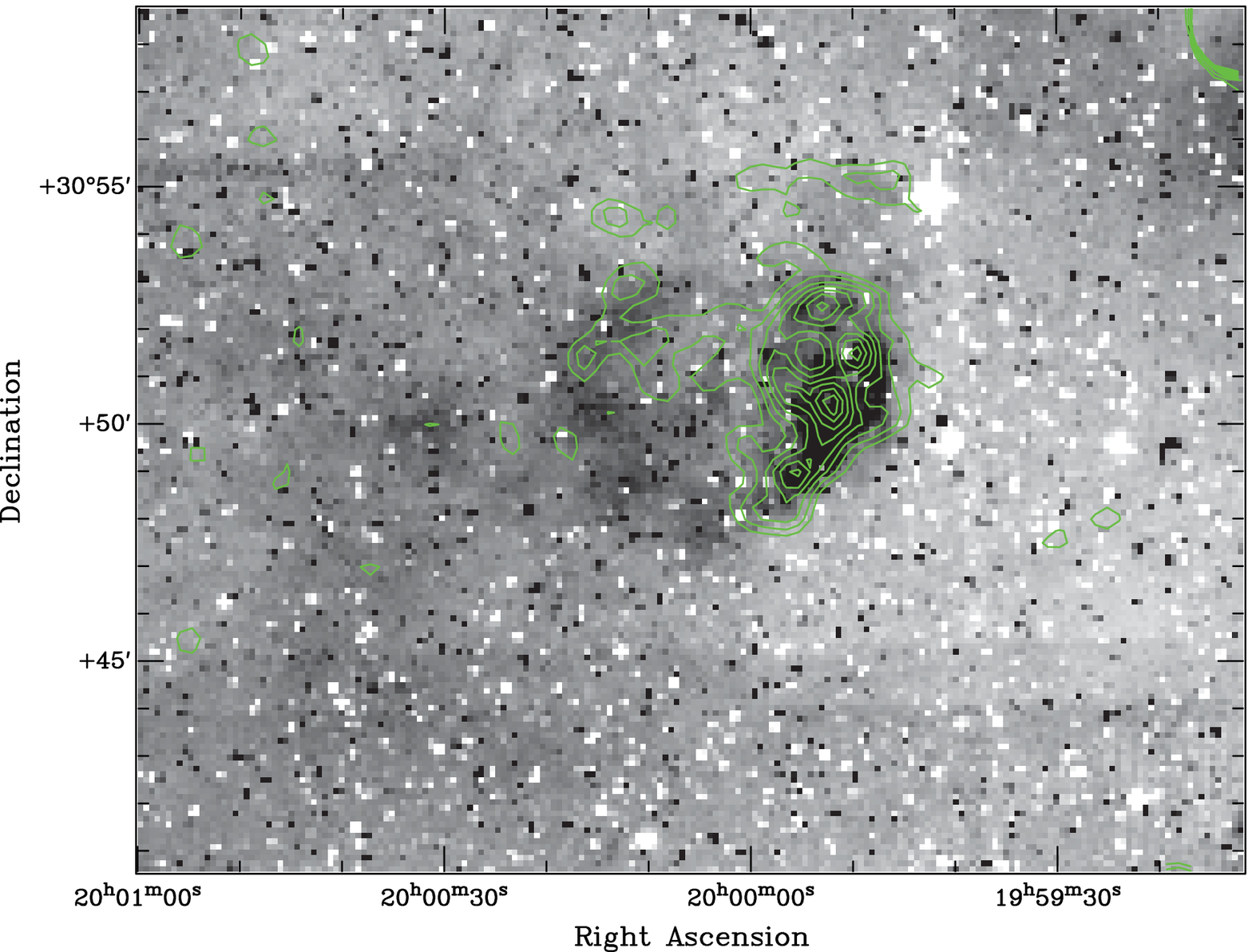}} 
{\includegraphics[height=5.4cm]{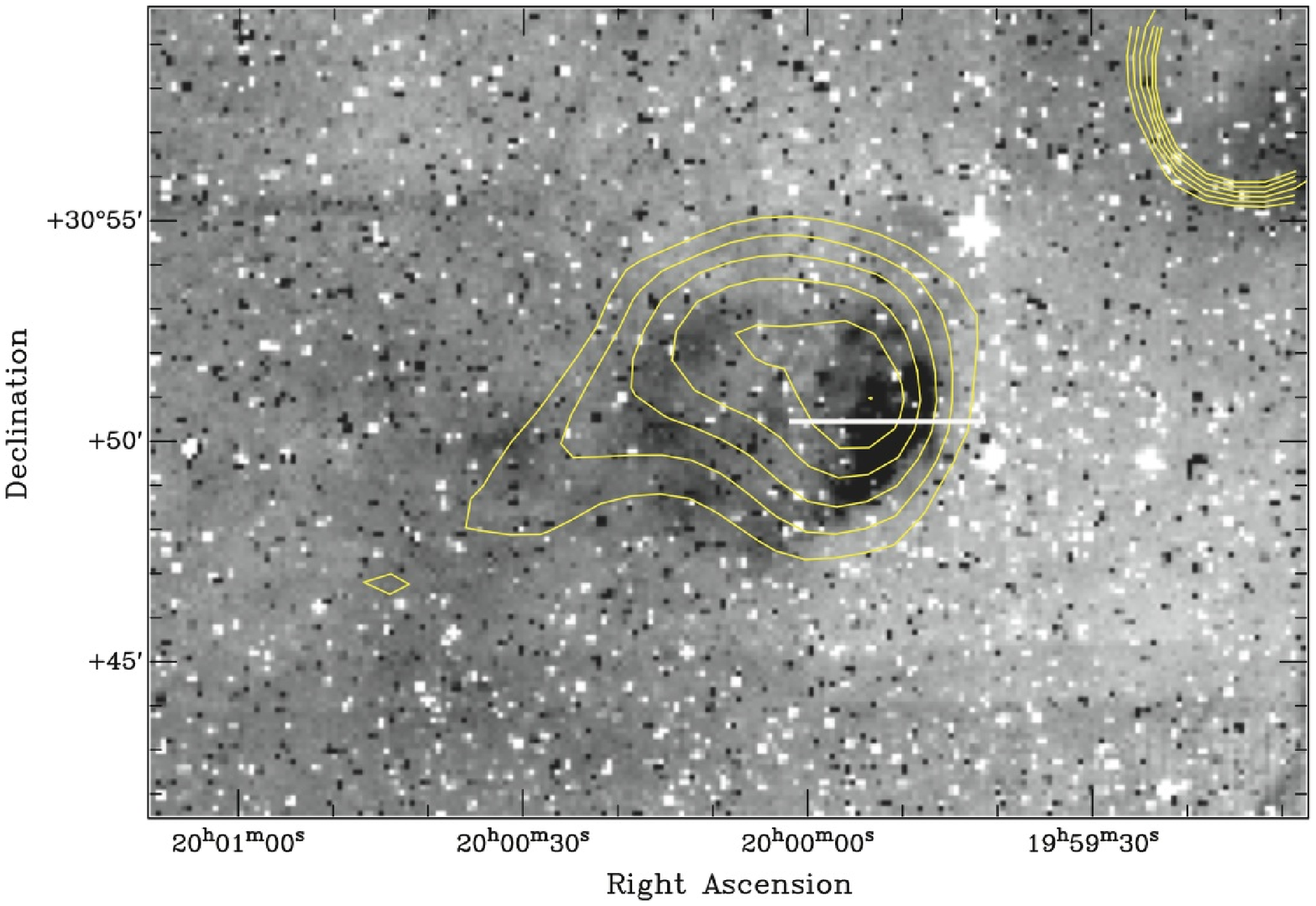}}
\hspace{-2cm}
\caption{\label{PointingD} Top panel:  The H$\alpha$-r image mosaic of IPHASX J200002.4+305035 with 15$\times$ pixel binning overlaid with NVSS contours in green ranging from 0.0015 to 0.004 Jy/beam. The E-W slit location of this `D' pointing was placed mid-way up  the Western enhanced edge of the shell. This detached nebula is at $\sim$5\arcmin\ East of IPHASX J195744.9+305306. We note that the NVSS 1.4~GHz emission is mainly coincident with the bright optical interacting front on the West. In the bottom panel the 87 GB 6~cm emission, overlaid as yellow contours ranging from 0.015 to 0.05 Jy/beam, is distributed over the whole nebula including out to the faint opposing optical emission rim directly to the West.  The E-W slit position is indicated.}
\end{figure}

\begin{figure}
\begin{center}
\hspace{-1cm}
{\includegraphics[height=8cm]{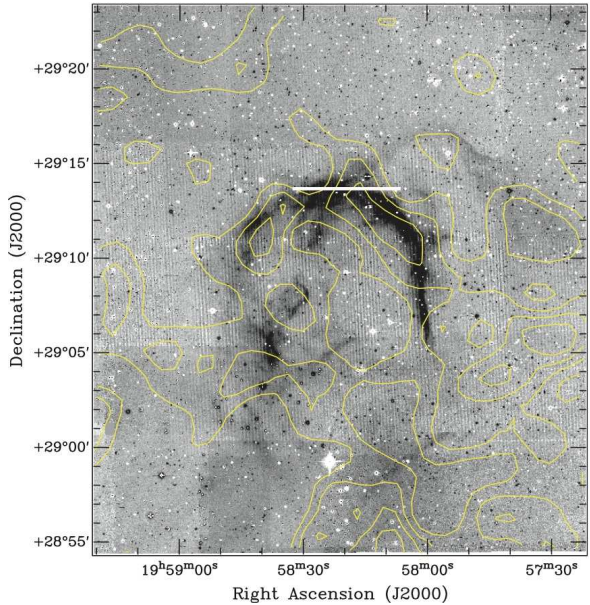}}
\caption{\label{195749} Image of new IPHAS SNR IPHASX J195749.2+290259. We show the optical IPHAS H$\alpha$ image overlaid with yellow 6~cm 87 GB contours which range from 0.0014 to 0.14 Jy/beam. There is a clear correspondence between the optical and radio emission particularly on the North-West rim.  In this case we did not find any clearly associated 1.4~GHz emission.}  Again the E-W slit position is marked.
\end{center}
\end{figure}

\begin{figure}
\begin{center}
{\includegraphics[height=6cm]{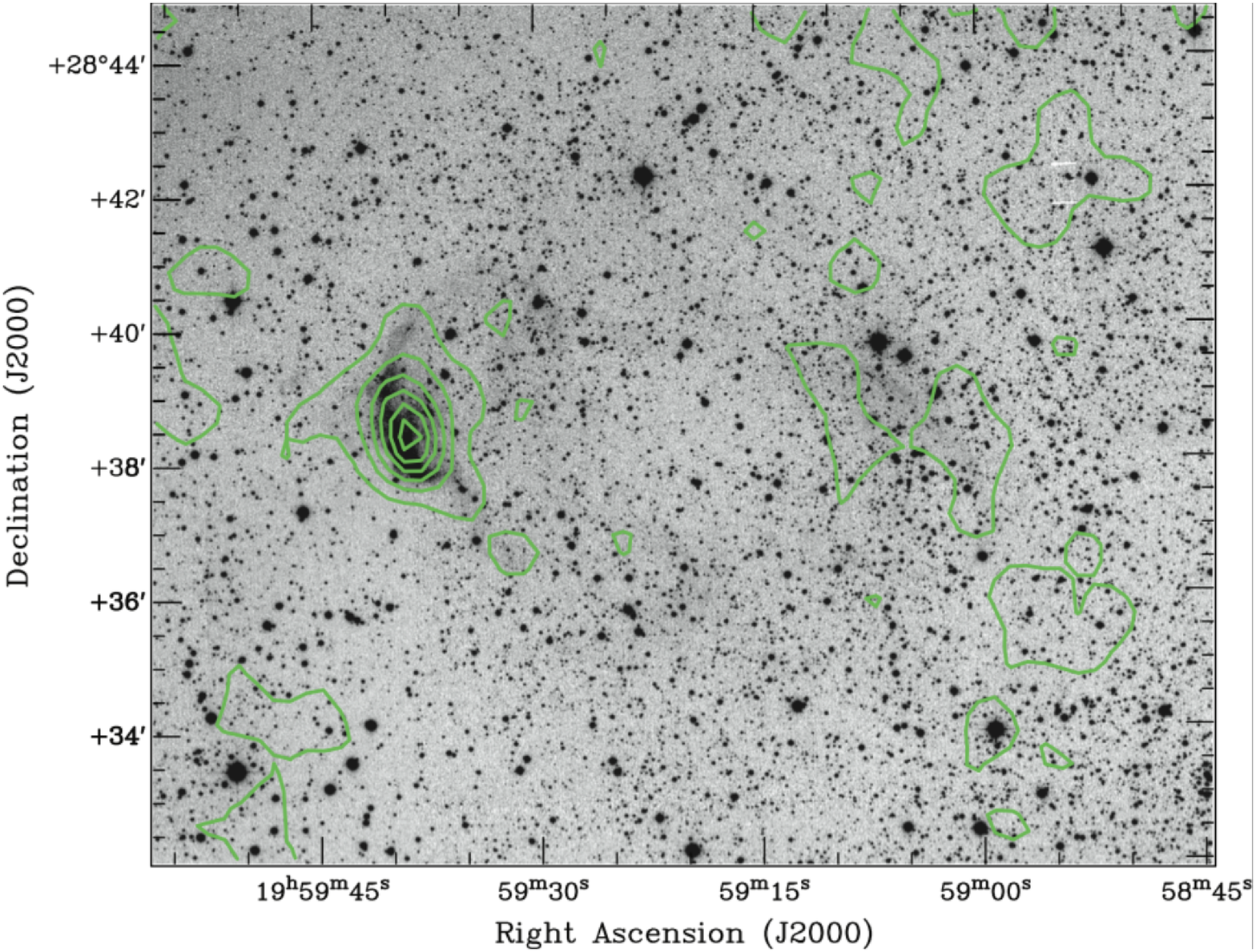}}
{\includegraphics[height=6cm]{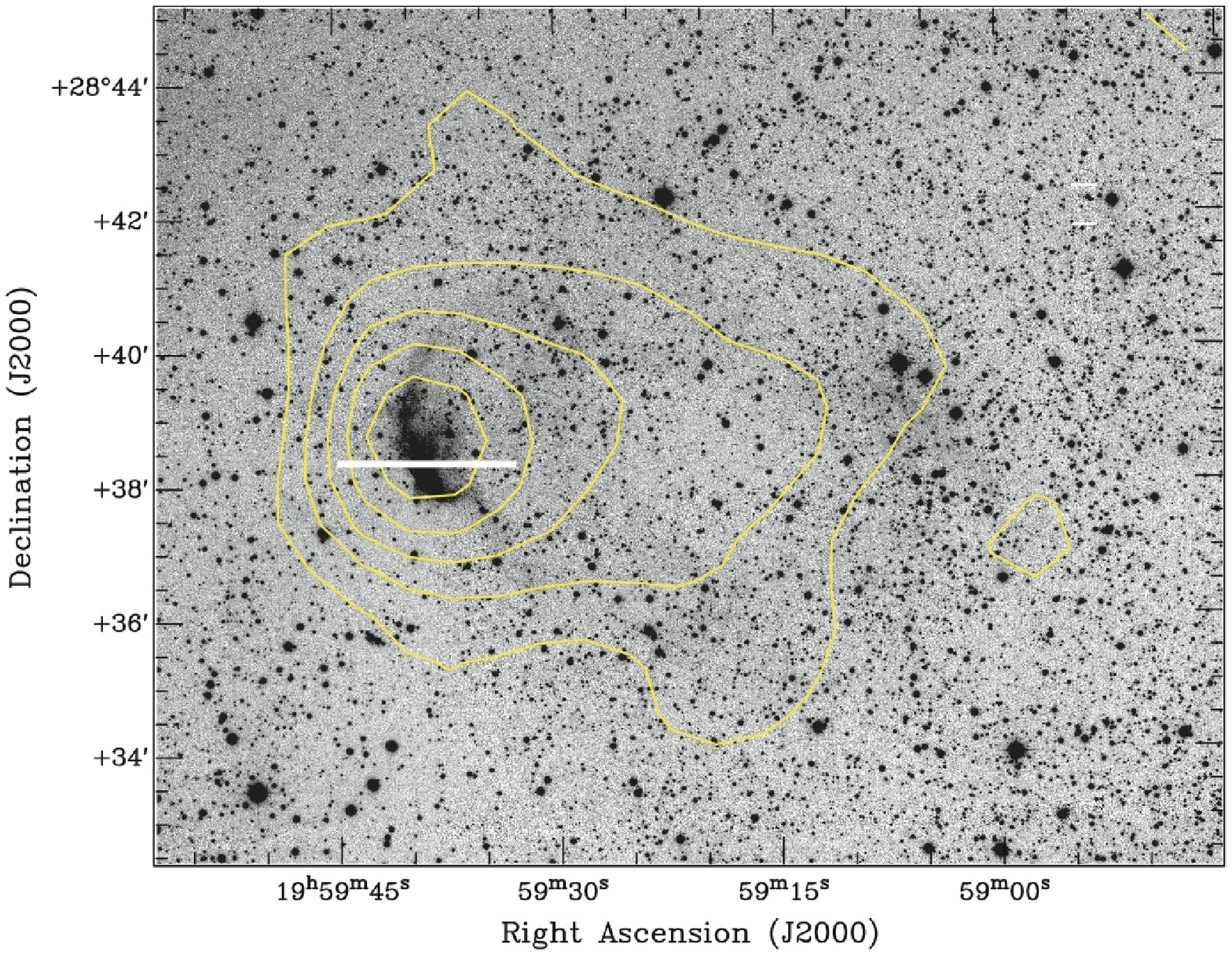}}
\caption{\label{SNR1959} Top panel: Full resolution H$\alpha$ mosaic of potential new SNR IPHASX J195920.4+283740 with the 21~cm NVSS (1.4~GHz) contours emission overlaid. The bottom panel, as above with now the  6~cm 87 GB radio image (4.85~GHz) contours emission overlaid. Those two maps indicates that while the eastern enhanced edge is mapped clearly by the 1.4~GHz emission, the 4.85~GHz radio emission is perfectly coincident with the full optical image which may partially reflect the lower resolution of the 87 GB data. A helium enriched emission line star is located close but internal to the eastern arc of the nebulae.  Spectrograph slit position marked E-W as before.}
\end{center}
\end{figure}

\section{Multi-wavelength data}

\subsection{G038.7-1.3 -- IPHASX J190640.5+042819} 
The region of sky containing this SNR was covered by two radio telescopes, the 64~m Parkes telescope as part of the PMN 6~cm survey 
\citep{Griffith1993} and the 91~m NRAO Green Bank telescope and included within the 87 GB survey \citep{Condon1989}.  Radio emission is clearly seen and follows quite well the optical nebular morphology as shown in the top panel of
Figure~\ref{SNR1906}. The yellow contours overlay the H$\alpha$ greyscale image and show the 4.85~GHz (6~cm) radio flux from the 87 GB survey as the resolution of this telescope at this frequency ($\sim$3.5\arcmin) is somewhat better than in PMN survey  ($\sim$5\arcmin). The radio data seems to indicate a much larger extent to the SNR than is evident from the optical data alone. Note that this radio emission had not been previously recognised as being part of any coherent structure so we make the association here for the first time. The radio contours go from 0.01 to 0.14 Jy/beam  in intervals of 0.02~Jy/beam.
  
The SNR is also seen at 1.4 GHz (21~cm) within the NVSS survey (Condon et al. 1998) and shown in the bottom panel of Figure~\ref{SNR1906}. Comparison with the 87 GB and PMN data shows that both the 6~cm and 21~cm surveys exhibit strong emission peak(s) at the same location (RA=19h~07m~20s, DEC=4$^{o}$ 32'). However, the NVSS image has much better resolution (45~arcseconds) and resolves two compact emission peaks at the same location. The radio contours go from 0.001 to 0.05 Jy/beam  in intervals of 0.005~Jy/beam. The integrated flux for this region provides values of 90~mJy for 87 GB and 109.2mJy and 119.9~mJy (21~cm) for the two NVSS frequencies available.
Unfortunately it is problematic to conclude if these more compact NVSS emission peaks are related to the SNR or are from a separate Galactic mJy source or even a background extragalactic double-lobed radio source.  Even if this double-peaked source is unrelated to the SNR the NVSS image still shows a complete, though fragmented SNR shell of $\sim$25\arcmin\ in diameter. Arrows are added to the figure to facilitate recognition. This enables an updated SNR ID of G038.7-1.3 based on the new geometric centre of the shell (note we retain the original IPHAS optical ID but the RA/DEC of the shell centre is approximately 19h~06m~35s +04$^{o}$~35'~00'').  Due to the low resolution at 6~cm  from the 87 GB survey we could not match emission with the NVSS shell at 21~cm apart from at the South-East edge. We also note that \citet{Reich1984} detected only one source (similar to the 6~cm 87 GB survey data) at the SNR location using the Effelsberg 100~m dish with a resolution $\sim4.3$\arcmin. 

 We emphasize, and this is also valid for the remaining radio analysis of the other objects, that measuring a valid spectral index for our SNR candidates (i.e. obtaining reliable radio fluxes at different frequencies) would be another clear proof to support their SNR status. However this was not possible with the existing data due to the non-compatibility between the radio surveys (different resolutions/beam sizes) and because of the generally fragmented nature of the remnants. New and deeper radio observations are required to obtain decent integrated radio fluxes from which reliable spectral indices could be obtained and hence to demonstrate the existence of non-thermal radio emission. This is beyond the scope of this paper.

A single X ray source is found within the confines of the SNR, ROSAT 1RXS~J190709.4+043100. Similarly to pulsars, a connection could be made via the localisation of X-ray sources in the central part of the SNR but this source is not located close to the geometric centre making any direct link unlikely. Furthermore, there is no indication of any extended X-ray emission internal to the shell which could indicate a Plerion. 

\subsection{G067.6+0.9 -- IPHASX J195744.9+305306}
The top panel in Figure~\ref{SNR195744.9} shows the H$\alpha$-r image mosaic of IPHASX J195744.9+305306 with 15$\times$ pixel binning overlaid with NVSS radio contours ranging from 0.0008 to 0.0027 Jy/beam. This higher resolution NVSS 21~cm emission directly and clearly overlaps with the strongest optical emission to the SSE. The complex environment surrounding the nebula makes it harder to define the real limits of the new SNR in 
H$\alpha$. The red E-W bars indicate the location of the slits for the different spectral pointings A, B and C with  lengths of 5\arcmin.  The bottom panel of Figure~\ref{SNR195744.9} is the same except with the 6~cm 87 GB radio emission overlaid as contours ranging from 0.004 to 0.02 Jy/beam. Here the radio emission appears coherently distributed along the Southern rim of IPHASX~J195744.9+305306. There are also some smaller, sparse emission blobs outside the nebula which are probably unrelated. Detection of  radio emission at two frequencies across components of the optical emission lends strong support to a likely SNR identification when combined with the indicative spectroscopic signatures. A group of X-ray sources has been identified within the boundaries of the proposed new SNR. Except for one located in the North-East (detected with the ROSAT all-sky survey catalogue of optically bright OB-type stars, \citealt{Berghoefer1996}) and related to the double/multiple star HR 7640, the majority of these X-ray sources are concentrated at $\sim$11.5\arcmin  West from the geometric centre of the SNR and are mostly ROSAT bright sources \citep{Voges1999} unrelated to the SNR.

\subsection{G067.8+0.5 -- IPHASX J200002.4+305035}
The possible new SNR PHASX~J200002.4+305035 has clear radio detections in both the NVSS and 87 GB. We present the NVSS data overlaid as green contours ranging from 0.0015 to 0.0040 Jy/beam on a grey-scale 15$\times$ pixel binned IPHAS 25$\times$20\arcmin\ H$\alpha$ image in the top panel of Figure~\ref{PointingD} due to its higher resolution. The bottom panel is as above with 87 GB data overlaid as yellow contours ranging from 0.015 to 0.05 Jy/beam. While the 6~cm 87 GB emission appears to entirely cover the optical nebula  including a weak extension to the Eastern side, the higher resolution  NVSS data is  more localized and coincides most closely with the Western enhancement of the oval shell (a possible bow-shock ?). 

\subsection{G066.0-0.0 -- IPHASX J195749.2+290259}
A multi-wavelength investigation of archival data reveals radio emission at 6cm from the 87BG survey. Figure~\ref{195749} shows the IPHAS H$\alpha$ grey-scale image overlaid with yellow 6~cm contours ranging  from 0.0014 to 0.14 Jy/beam that appears to follow the edges of the SNR especially to the North--East and North--West though it is also somewhat fragmented.  An interesting point is the location only $\simeq$ 31\arcmin to the South--East of another of our new IPHAS identified SNRs G065.8-0.5 (IPHASX J195920.4+283740) further discussed below.  The E-W slit position of the spectroscopic observation is indicated by the white horizontal bar.

\subsection{G065.8-0.5 -- IPHASX J195920.4+283740}
For this SNR candidate an excellent match between the known NVSS radio source NVSS~195938+283832 (27.0~mJy at 1.4~GHz) and the bright Eastern enhanced arcuate rim of this possible bow shock was found  (see top panel of Figure~\ref{SNR1959}). 
There is also a very good coincidence between the 4.85~GHz 6~cm 87 GB radio emission and the entire optical nebula extent including the fainter opposing rim to the West (see bottom panel of Figure~\ref{SNR1959}). When taken together with the optical emission line ratios, which place the nebula in the SNR regime of the various diagnostic diagrams, an SNR identification seems likely. However, the situation is complicated by possible association with a helium enriched emission-line star \citep{Corradi2010} close to but internal to the Eastern arc traced by the optical and NVSS data. More investigation is  needed to  ascertain the true nature of this nebula.

\begin{figure*}
\begin{center}
{\includegraphics[height=17cm]{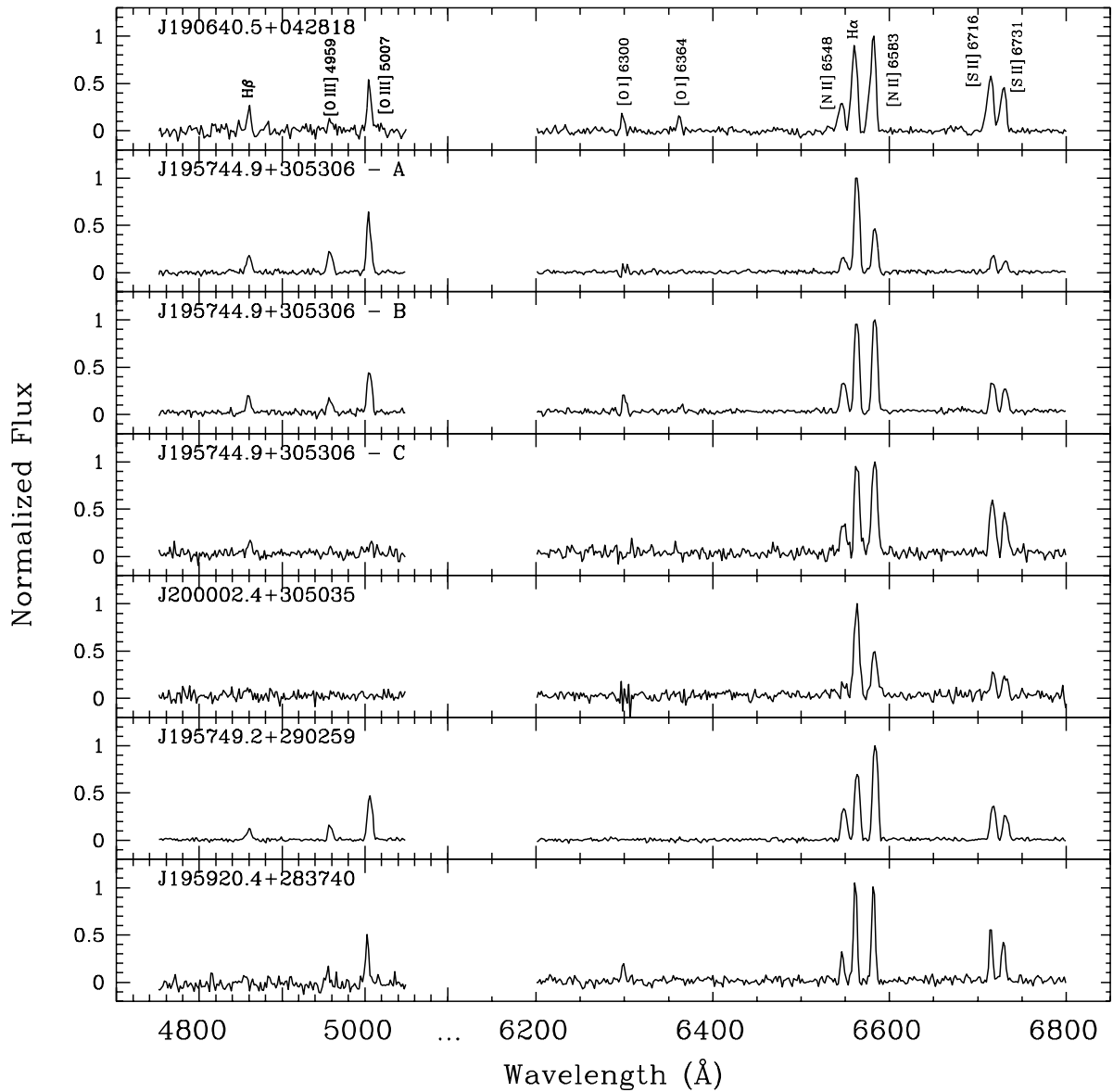}}
\caption{\label{Spectra} Blue and red spectral regions of interest for the five IPHAS targets. The main emission lines are labelled and the fluxes are normalized. The spectra have been cropped for a better display of the nebular lines but the area of the gap is empty of lines or are inside the noise level. We recall that IPHASX J200002.4+305035 also designates the pointing D.}
\end{center}
\end{figure*}

\begin{figure*}
\begin{center}
\includegraphics[height=10.25cm]{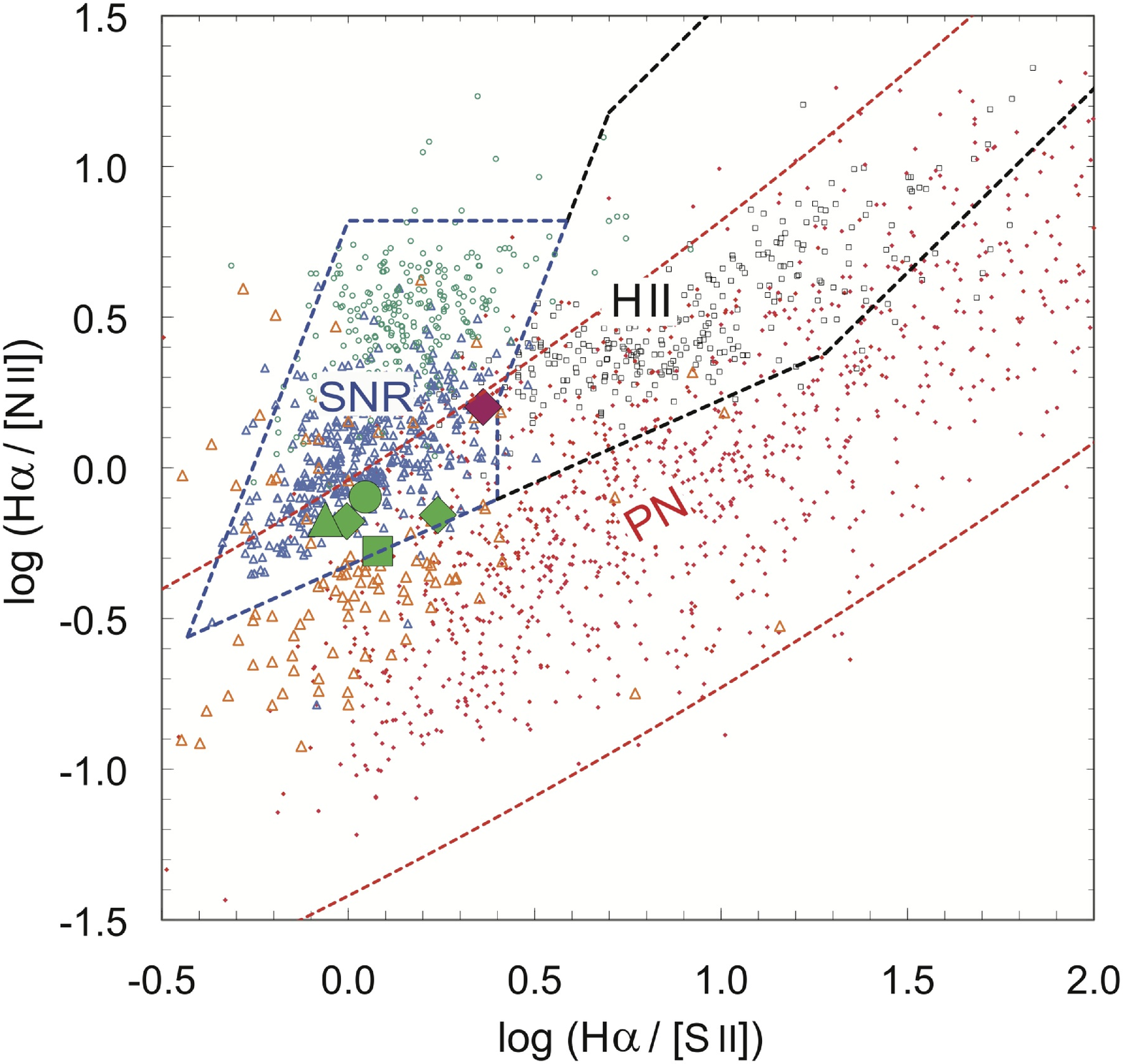}
\includegraphics[height=10.25cm]{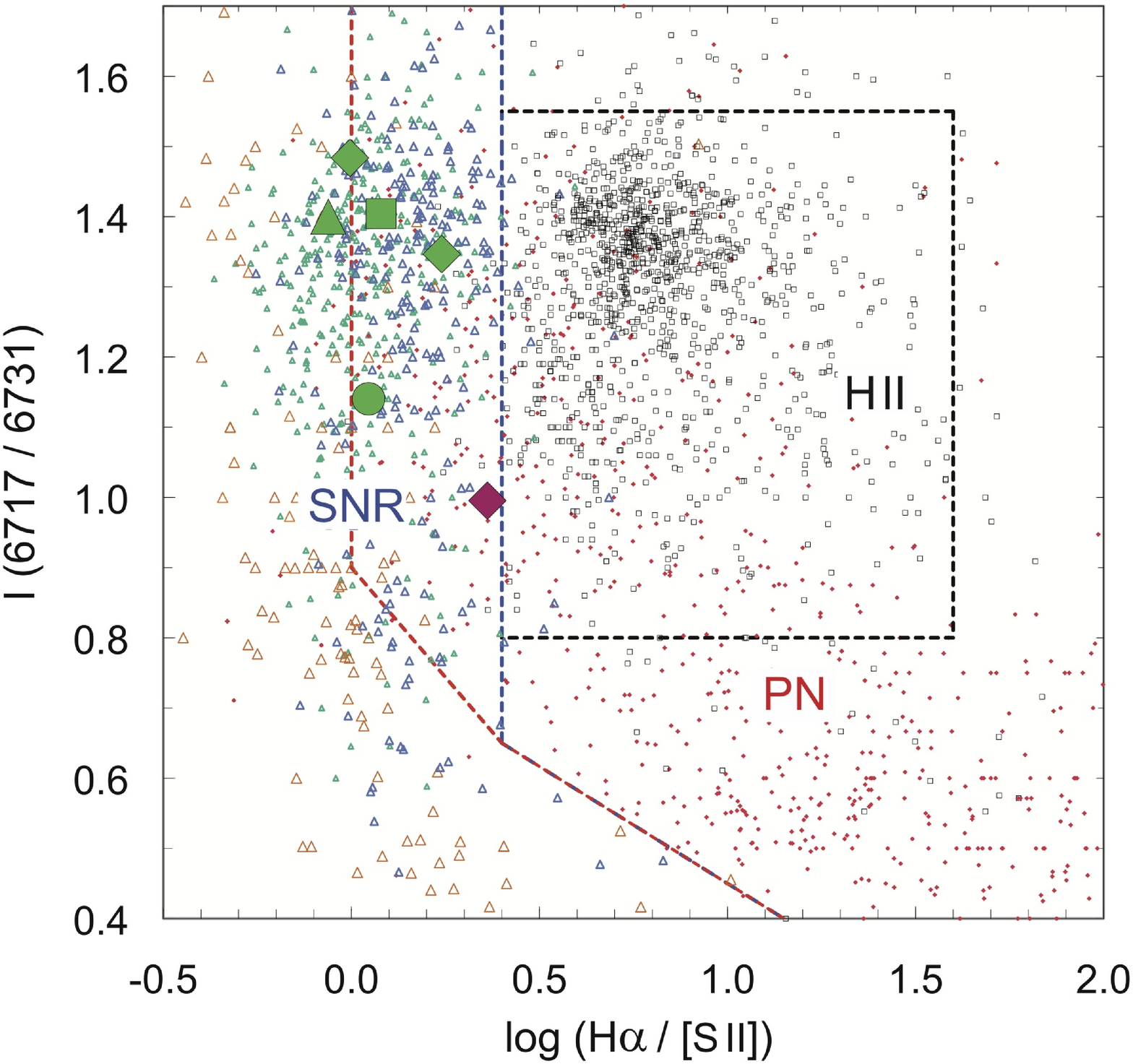}
\caption{\label{DJF}Top panel: The  `SMB' H$\alpha$/{[N \textsc{ii}]} versus H$\alpha$/{[S \textsc{ii}]} diagnostic diagram updated from \citep{Frew10} (refer to the text). The five new IPHAS sources are plotted showing their likely SNR nature: IPHASX J19640.5+042819 (large green triangle), IPHASX J195749.2+290259 (green square), IPHASX J195920.4+283740 (green circle), IPHASX J195744.9+305306 for which several spectra were obtained (Figure~\ref{SNR195744.9}) and for which we plot pointings B and C (green diamonds), and IPHASX J200002.4+305035 (magenta diamond) which is pointing~D in Figure~\ref{PointingD}. Galactic PNe are plotted as small red dots, HII regions as small open black squares, evolved Galactic SNRs as small open blue triangles, young Galactic SNRs as orange triangles, and the low-metallicity Magellanic Cloud SNRs as small green triangles. The PN domain is bounded by the red dashed curves, the evolved SNR field by the dashed blue lines, and the HII region domain by the dashed black lines.  Note the considerable overlap between the PN and HII region fields.
Lower panel: The SMB H$\alpha$/{[S \textsc{ii}]}$\lambda$6717/$\lambda$6731 versus 
H$\alpha$/{[S \textsc{ii}]} diagnostic diagram. The domains and symbols are defined as in the top hand panel.  All new IPHAS nebulae fall within the SNR fields on both plots lending considerable weight to their SNR identification.  A colour version of this figure is available in the online journal.}
\end{center}
\end{figure*}

\begin{figure*}
\begin{center}
\hspace{-2.8cm}
\includegraphics[height=10.5cm]{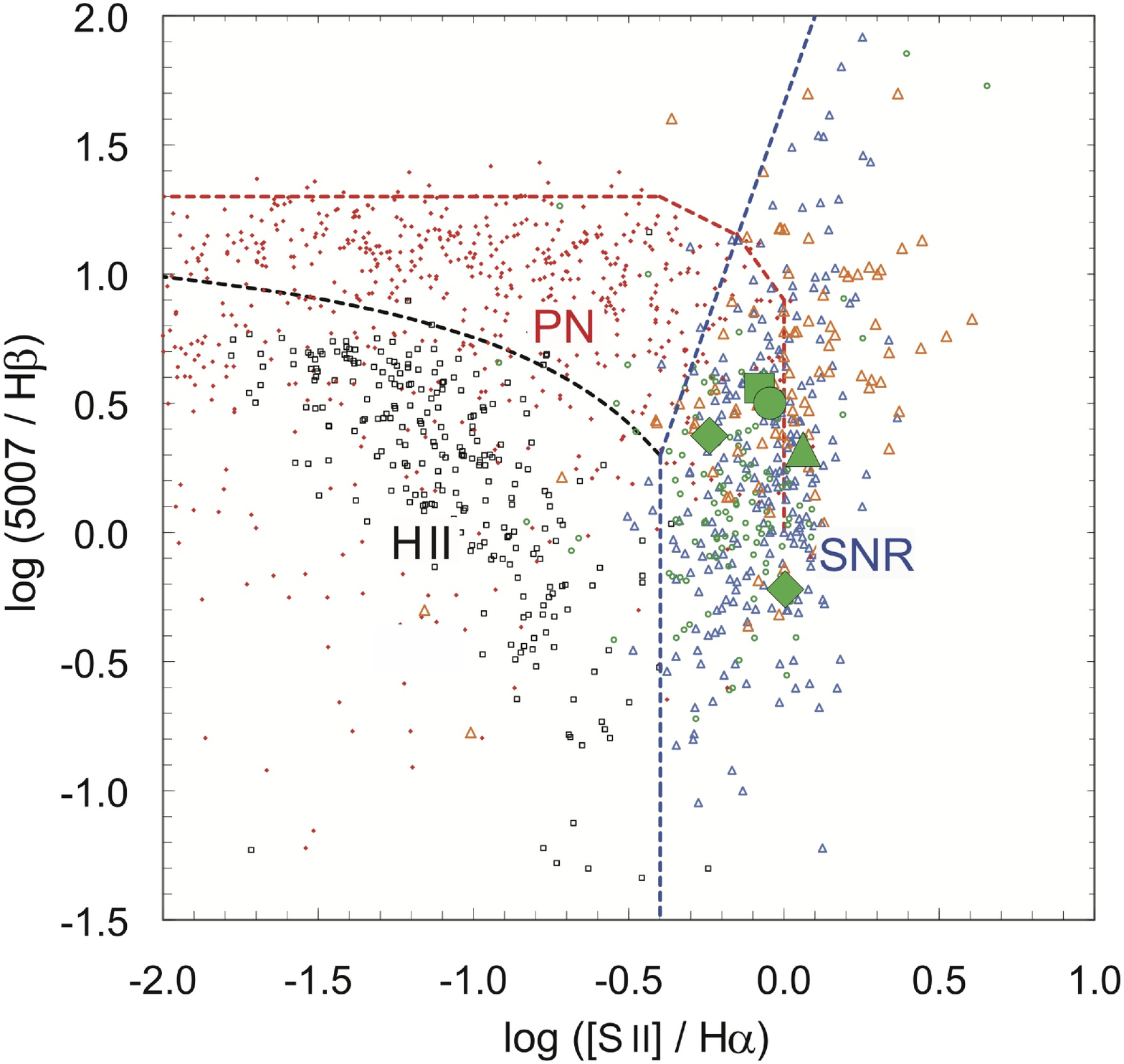}
\includegraphics[height=10.5cm]{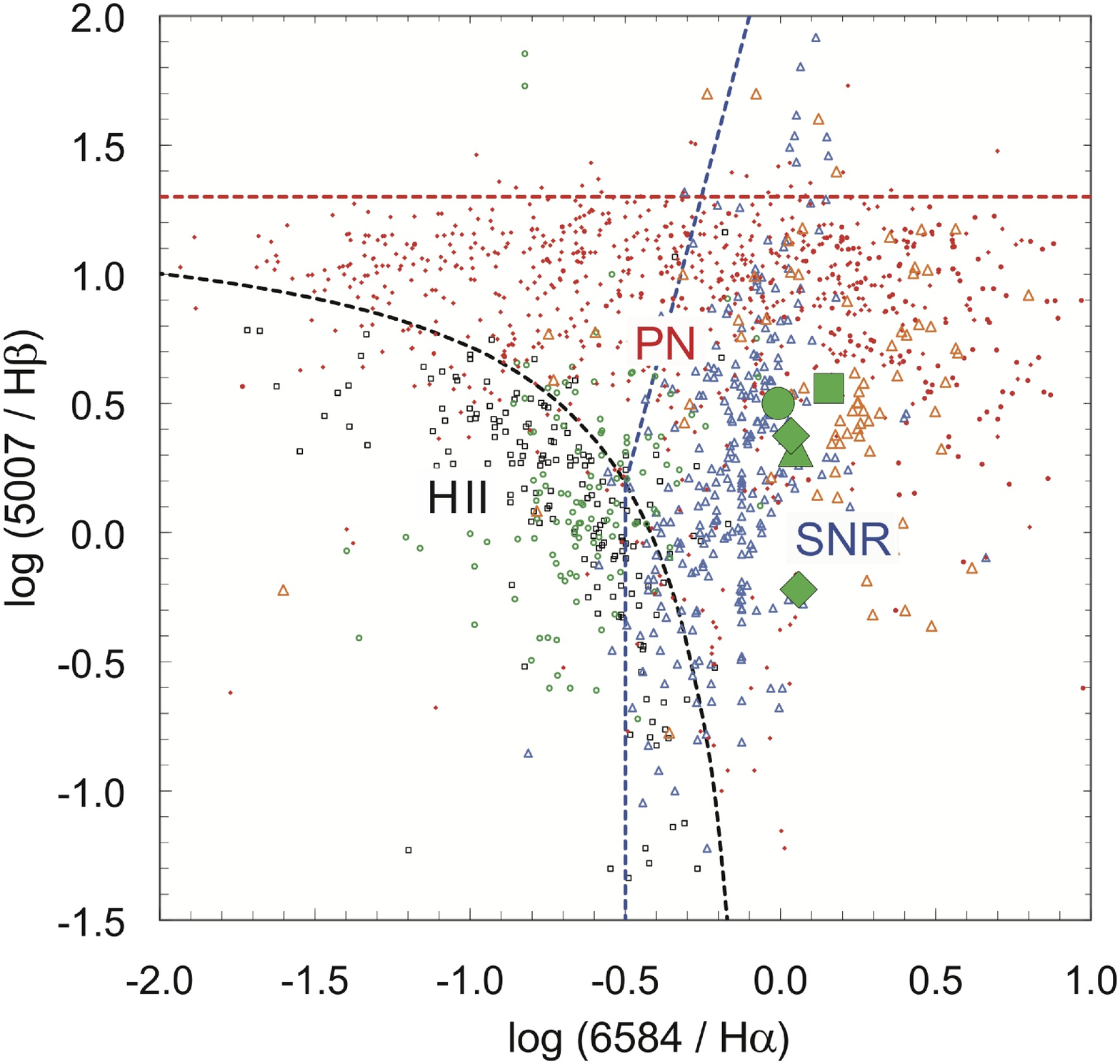}
\caption{\label{DJF2}Top panel: The `BPT' {[O\,\textsc{iii}]}$\lambda$5007/H$\beta$ versus {[S \textsc{ii}]}/H$\alpha$ diagnostic diagram with the domains defined as before. The symbols are the same as in Figure~\ref{DJF}.  Lower panel:  The BPT {[O\,\textsc{iii}]}$\lambda$5007/H$\beta$ versus {[N \textsc{ii}]}$\lambda$6584/H$\alpha$ diagram.   As in Figure~\ref{DJF}, four of the new IPHAS nebulae fall well within the SNR domains in both plots. IPHASX J200002.4+305035 is not plotted owing to our non-detection of [O \textsc{iii}] emission in this case.  A colour version of this figure is available in the online journal.}
\end{center}
\end{figure*}


\begin{table*}
\caption{\label{Em} {Relative observed emission line fluxes and flux ratios obtained (F($\lambda$)) and corrected for extinction (I($\lambda$)) for the IPHAS sources with respect to H$\beta$=100. The numbers in parenthesis indicate the signal-to-noise ratios.}}
\resizebox{1.12\linewidth}{!} {
\begin{tabular}{|l|l|l|l|l|l|l|l|l|l|l|l|l|c|c|l|l|l|l|l|l|l|} 
\hline
&  \multicolumn{2}{|c|}{J190640.5+042819}&  & \multicolumn{9}{|c|}{J195744.9+305306}& \multicolumn{1}{|c|}{J200002.4+305035} & & & \multicolumn{2}{|c|}{J195749.2+290259} & &   \multicolumn{2}{|c|}{J195920.4+283740}\\  \cline{2-3}\cline{5-12}\cline{14-15}\cline{17-18}\cline{20-21}

&   \multicolumn{2}{|l|}{}  &  & \multicolumn{2}{|c|}{A} &  & \multicolumn{2}{|c|}{B}&  & \multicolumn{2}{|c|}{C} & & 
 \multicolumn{1}{|c|}{D} &   \multicolumn{2}{|c|}{} &   \multicolumn{2}{|c|}{}  \\  \cline{5-6}\cline{8-9}\cline{11-12}

 ID, \AA &  F($\lambda$) & I($\lambda$) &  &F($\lambda$) & I($\lambda$)&  &F($\lambda$) & I($\lambda$)& & F($\lambda$) & I($\lambda$)&  &F($\lambda$) &  & & F($\lambda$) & I($\lambda$)& &  F($\lambda$) & I($\lambda$) \\
\hline 
H$\beta$ 4861         & 100 (1.7) & 100  && 100(3.3)& 100  && 100(2.8)  & 100  && 100(2.2)  & 100  &&--   &&& 100(3.7)  & 100   && 100(1.1$\dagger$)  & 100  \\
{[O\,{\sc iii}]} 4959 & 53.9(1.4) & 51.5   && 124.4(3.0)& 118.6&& 86.4(2.3) &82.2  &&38.2($\sim$1$\dagger$) &35.5  && --  &&&118.3(3.2)  &112.1  && 86.3($\sim$1$\dagger$) & 80.9  \\
{[O\,{\sc iii}]} 5007 & 224.8(2.1)& 210.3&& 274.5(3.6)& 255.9&& 254.9(5.5)&236.8 &&67.1($\sim$1$\dagger$) &60.2  && --  &&&392.2(3.3)  &361.9 && 348.0(2.0)& 316.4  \\
{[O\,{\sc i}]} 6300   & 59.5(1.5) & 34.1 && --   & --               && 76.0(2.7) &41.2 && --   &--    && --   &&& --   &--    && 117.1(1.7)& 52.9  \\ 
{[O\,{\sc i}]} 6364   & 61.4(2.0) & 34.5   && --   & --                && 21.4(1.8) &11.4  && --   &--    && --    &&& --   &--    && --   &  -- \\
{[N\,{\sc ii}]} 6548  & 194.8(1.2)& 103.6&& 88.4(1.0) & 45.6  && 201.9(1.3)&100.8&&284.5(1.2)&102.8 && 9.9(1.1) &&&280.9(3.2)  &131.5 &&187.5(1.9) &76.2   \\
H$\alpha$ 6563        & 539.4(2.8)& 286.0&& 556.8(4.7)& 286.0 && 575.4(3.3)&286.0&&796.3(2.1)&286.0 && 100(3.2) &&&613.8(3.1)  &286.0  &&707.5(5.2) &286.0   \\ 
{[N\,{\sc ii}]} 6583  & 604.4(2.9)& 318.7&& 255.5(2.0)& 130.5 &&625.8(3.6)&309.2&&918.8(2.2)&327.1 && 52.9(1.6)&&&883.0(3.3)  &408.7 &&699.3(5.1) &280.5   \\
{[S\,{\sc ii}]} 6716  & 378.4(2.3)& 192.2 && 85.4(2.3) & 41.9&& 199.5(2.8)&94.6&&511.3(2.2)&171.2&& 21.7(2.2)&&&315.0(2.9)  &139.3  &&361.0(2.8) &137.2   \\
{[S\,{\sc ii}]} 6731  & 271.4(2.0)& 137.3 && 57.6(1.6) & 28.2&& 148.4(2.0)&70.1&&339.8(1.6)&113.1&& 21.8(1.9)&&&225.3(2.5)  &99.2 &&318.3(2.3) &120.3  \\
\hline
 Log[F(H$\beta$)]& -12.17 &&& -11.65  &&& -11.62  &&&-13.08 &&& -- &&&-12.72 &&&-15.78 \\
 
 $c$(H$\beta$)& 0.78$\pm$0.11 &&& 0.82$\pm$0.12&&& 0.86$\pm$0.11 &&&  1.26$\pm$0.10&&&-- &&&0.94$\pm$0.11&&&1.12$\pm$0.16   \\
 Av          & 1.69$\pm$0.24 &&& 1.77$\pm$0.25&&& 1.86$\pm$0.24 &&&  2.73$\pm$0.23 &&&-- &&&2.03$\pm$0.24&&&2.42$\pm$0.35   \\ 
 \hline
([S\,{\sc ii}]/H$\alpha$)$_{c}$& 1.15$\pm$0.14&&& 0.24$\pm$0.12 &&&0.57$\pm$0.13&&&0.99$\pm$0.13 &&& 0.44$\pm$0.12 &&& 0.83$\pm$0.13 
&&&  0.90$\pm$0.16  \\
([N\,{\sc ii}]/H$\alpha$)$_{c}$& 1.48$\pm$0.16&&& 0.62$\pm$0.14 &&&1.43$\pm$0.16&&&1.50$\pm$0.15 &&&0.63$\pm$0.13 &&&1.89$\pm$0.18 
 &&& 1.25$\pm$0.21  \\
(6716/6731)$_{c}$  & 1.40$\pm$0.11&&   & 1.49$\pm$0.15 &&  &1.35$\pm$0.1&&&1.51$\pm$0.08 &&&0.99$\pm$0.04 &&&1.40$\pm$0.11  
&&& 1.14$\pm$0.17  \\
\hline
\end{tabular} }
\begin{minipage}[b]{1\linewidth}
Note: (*)$_{c}$ indicates the ratios with values corrected for extinction.\\
$\dagger$ Very low S/N leading to high uncertainty on the line measurement.\\
Pointing D: Due to the absence of H$\beta$, the fluxes are set respective to H$\alpha$ = 100.\\
{[S\,{\sc ii}]}/H$\alpha$ = ([S\,{\sc ii}]6716 + [S\,{\sc ii}]6731) / H$\alpha$ \\
{[N\,{\sc ii}]}/H$\alpha$ = ([N\,{\sc ii}]6548 + [N\,{\sc ii}]6583) / H$\alpha$ \\
Log F(H$\beta$) units: erg s$^{-1}$ cm$^{-2}$ \AA$^{-1}$\\
\end{minipage}
\end{table*}

\section{Discussion and Conclusions}

 We report the discovery of four or possibly five new, likely SNRs following a systematic and careful visual search of the continuum removed H$\alpha$ mosaics of the IPHAS survey over the 19-20~hour RA zone. These small-to-medium angular size Galactic nebulae and SNR candidates  (i.e. with major axes ranging from  $\sim$6--50\arcmin) were selected primarily on the basis of their optical morphology. Plausible candidate SNRs need to exhibit coherent filamentary or shell-like characteristics which are common features of currently known optically identified SNRs. The relatively small angular size of these objects compared with the degree-size SNRs found for many Galactic remnants  (e.g. refer to on-line catalogue of Green -- http://www.mrao.cam.ac.uk/surveys/snrs/snrs.data.html) implies that these newly discovered nebulae  could be more distant or perhaps younger, assuming they have not had the time to expand to the dissipation phase. In the latter case a connection between an SNR and a young pulsar might be realised if they originate from core-collapse events.  

 Our spectroscopic follow-up of these candidates allowed us to identify them all as likely SNRs using the standard diagnostic emission line ratios defined by \citet{Fesen1985} and the new and more comprehensive diagrams presented first by \citet{Frew10} and with updated and improved versions of these figures presented here for the first time. A problem in this type of investigation lies in the generally faint optical emission and low surface brightness of such evolved SNRs which makes their spectroscopic investigation difficult on 2~m class telescopes. Nevertheless, in this preliminary study, the primary combination of optical imagery and spectroscopy  and their locations in the emission line diagnostic plots lend strong support to the SNR origin of the examined candidates. In several cases this has been further corroborated by the recognition of previously un-catalogued radio emission  usually at more than one frequency  from archival data which is clearly associated with the optical nebulosities. 

The spectroscopic data show typical SNR diagnostic lines exhibited by all candidates and indicates the presence of shocks with  [S\,{\sc ii}]/H$\alpha$ line ratios greater than 0.4--0.5. The observed  [N\,{\sc ii}]/H$\alpha$ line ratios are also generally high ($>1.2$) which rule out any possible confusion with HII regions. There are two exceptions. The first is for pointing `A' of IPHASX J195744.9+305306 which has  [N\,{\sc ii}]/H$\alpha$~$\sim$0.62. However, we determine that this compact nebular region, which is located internal to the main nebula shell and targeted as pointing `A'  is likely to be an unrelated HII region. The second is IPHASX J2000002.4+305035. In this case the spectrum still displays an  [N\,{\sc ii}]/H$\alpha$ $>0.5$ which is atypical of HII regions. The [OI] 6300 and 
6364\AA\ emission often seen in optical remnants are difficult to unambiguously identify in most cases due to low spectral resolutions employed meaning that these lines cannot be easily resolved or subtracted properly from the strong night sky lines of the same species due to the lack of significant line velocity difference. Consequently, among the group of likely SNRs, only three show [OI]~6300\AA\ and/or [OI]~6364\AA~ with any certainty though the line intensities themselves are only indicative and are not recorded in the canonical 3:1 ratio. 

It is also important to emphasise that  shock emission is not inherent solely to SNRs and can result from stellar winds and interactions such as those produced by young-stellar objects and some PNe and WR stars. However, in general WR shells do not show strong  [S\,{\sc ii}] emission, and largely plot in separate regions of \citet*{Sabbadin77} (SMB) type diagnostic plots.

In order to further corroborate our results we also performed a multi-wavelength investigation based on searches of online archives including all available radio  data from the various current on-line surveys. Radio emission, mainly at 4.85~GHz (6~cm) from the 87 GB  radio survey, was found in all cases with a decent positional and structural match to components of the optical morphology. In most cases there were also detections from the higher resolution 1.44~GHz (21~cm) NVSS survey data which provides clearer and more direct associations. These radio matches are particularly good in the cases of G038.7-1.3 (IPHASX J190640.5+042819), G065.8-0.5 (IPHASX J195920.4+283740) and G067.8+0.5 (IPHASX J200002.4+305035) which supports the SNR nature of those objects. The remaining two SNRs display more discontinuous radio emission but still with components that follow the most intense regions of optical emission.  Full overlap between the radio and optical data is not a necessary condition to demonstrate association, particularly in the case of faint nebulae and possible local dust and density variations can affect the visibility of optical emission across the true extent of the SNR.

Another source of confusion with possible SNR identifications of optical nebulae resides in the diffuse ionised gas (DIG) emission (or warm interstellar medium, WIM) as its morphology can be confused with our equally faint SNR candidates. Since the DIG is generally photo-ionised, as are HII regions \citep{Domgorgen1994}, the separation of the DIG with shock-ionised objects such as SNRs can be easily done from our optical spectroscopy using the limiting conditions for both states:  [S\,{\sc ii}]$\lambda${$\lambda$}6716,6731/H$\alpha$ $\simeq$ 0.4 and  [N\,{\sc ii}]$\lambda${$\lambda$}6548,6583/H$\alpha$ $\simeq$ 0.5 (see for example \citealt{Galarza1999}, their figure 8). As mentioned previously, our line-ratios clearly indicate that except for region `A' in IPHASX J195744.9+305306 (which is indeed likely an unrelated HII region) and the ambiguous IPHASX J2000002.4+305035, all the targets show shock-ionisation characteristics of SNRs ( [S\,{\sc ii}]/H$\alpha$ $>$ $\sim$0.4 and  [N\,{\sc ii}]/H$\alpha$ $>$ $\sim$0.5. A DIG classification can therefore be confidently discarded for these sources.

 Figure \ref{Spectra} and Table \ref{Em} indicate a relatively strong [OIII]5007\AA~which is known to be a good indicator of higher shock speed.  \citet{Raymond1979} showed that the presence of this line is coincident with shock speed greater than 80~km/s. This trend is confirmed when comparing our results with the theoretical models by \citet{Allen2008} which indicate  shock speeds between 200 and 1000 km/s in the SNR where [OIII]5007\AA is ``well" detected. This does not apply to the C slit-position of J195744+305306 where the [OIII] 5007\AA~line is hardly detectable and most probably mixed with noise. Those high values would help to constrain the evolutionary phase of the SNR observed, as although we are not dealing with SNRs in their ultimate phase of dissipation in the ISM we are not dealing with young ones either.

Our results on the first SNRs identified from IPHAS are preliminary, and each nebula warrants additional follow-up.  Our ongoing search for new SNR candidates is based on careful and systematic examination of the IPHAS survey image mosaics of the Galactic plane and is highly likely to provide many more SNR identifications in the future.  Considering the return from a single 1-hour IPHAS RA zone we might expect dozens of new Galactic SNR candidates to be uncovered based on their optical emission signatures.  This would add significantly to the known Galactic population of evolved SNRs.

\section*{Acknowledgements}
We thank both the anonymous referee and P. Boumis for valuable comments and suggestions that have contributed to significantly improving the quality of the paper. This project has been supported by PAPIIT-UNAM grant IN109509. This paper makes use of data obtained as part of the INT Photometric H$\alpha$ Survey of the northern Galactic plane (IPHAS) carried out at the Isaac Newton Telescope (INT). The INT is operated on the island of La Palma by the Isaac Newton Group at the Spanish Observatorio del Roque de los Muchachos of the Instituto de Astrofisica de Canarias. All IPHAS data are processed by the Cambridge Astronomical Survey Unit, at the Institute of Astronomy in Cambridge. The investigation is based on observations collected at the 2.1 m telescope at the Observatorio Astron\'{o}mico Nacional at San Pedro M\'{a}rtir (Baja California, M\'{e}xico) operated by the Instituto de Astronom\'{i}a of the Universidad Nacional Aut\'{o}noma de M\'{e}xico and at the 2.5m INT telescope operated on the island of La Palma by the Isaac Newton Group in the Spanish Observatorio del Roque de los Muchachos of the Instituto de Astrof\'{i}sica de Canarias. Special thanks are given to the technical staff and night assistants of the San Pedro M\'{a}rtir Observatory. This research has made use of the ALADIN database operated at the CDS, Strasbourg (France) and the NASA/ IPAC Infrared Science Archive, which is operated by the Jet Propulsion Laboratory, California Institute of Technology, under contract with the National Aeronautics and Space Administration. This research also made use of Montage, funded by the National Aeronautics and Space Administration's Earth Science Technology Office, Computation Technologies Project, under Cooperative Agreement Number NCC5-626 between NASA and the California Institute of Technology. Montage is maintained by the NASA/IPAC Infrared Science Archive. MS thanks the ARC for provision of Discovery project funding while DJF thanks Macquarie University for a university research fellowship.

\bibliographystyle{mn2e}

\bibliography{sabin_snr}

\bsp
\label{lastpage}

\end{document}